\documentclass[11pt]{article}
\usepackage{cite}
\usepackage{amsmath,amsfonts,amssymb}
\usepackage[small,bf,hang]{caption}
\usepackage[dvips]{graphicx,epsfig}
\usepackage{slashed}
\usepackage[vcentermath]{youngtab}

\def\hybrid{
        \topmargin -20pt
        \oddsidemargin 0pt
        \headheight 0pt \headsep 0pt
        \textwidth 6.25in % A4 paper
        \textheight 9.5in % A4 paper
        \marginparwidth .875in
        \parskip 5pt plus 1pt \jot = 1.5ex}

\hybrid
\usepackage{tikz}

 \csname
@addtoreset\endcsname{equation}{section}

\usepackage{color}
\definecolor{red}{rgb}{1,0,0}
\definecolor{lred}{rgb}{0.3,0,0}
\definecolor{green}{rgb}{0,0.6,0}
\definecolor{blue}{rgb}{0,0,1}
\definecolor{violet}{rgb}{0.8,0,0.8}
\definecolor{amber}{rgb}{1.0, 0.75, 0.0}
\definecolor{yellow}{rgb}{1.0, 1.0, 0.0}
\definecolor{applegreen}{rgb}{0.55, 0.71, 0.0}
\definecolor{cadmiumgreen}{rgb}{0.0, 0.42, 0.24}
\definecolor{ballblue}{rgb}{0.13, 0.67, 0.8}
\definecolor{caribbeangreen}{rgb}{0.0, 0.8, 0.6}
\definecolor{bluemunsell}{rgb}{0.0, 0.5, 0.69}
\definecolor{brightpink}{rgb}{1.0, 0.0, 0.5}

\newcommand{\co}[1]{\text{csch}^{#1} \bar x}

\newcommand{\D}{{D}}
\newcommand{\hD}{\bar{{D}}}
\newcommand{\hM}{\bar{M}}

\newcommand{\nn}{\nonumber}

\newcommand{\tr}[1]{\text{Tr}(#1)}
\newcommand{\Tr}[1]{\text{Tr}\left(#1\right)}

\def\moth{\mathsurround=0pt}
\newdimen\zo \zo=0pt

\def\tick{\leaders\hrule height 0.5ex depth 0pt \hskip 0.5pt}
\def\upboxfill{$\moth \setbox\zo\hbox{\tick}%
  \hskip 3pt\hbox to 0pt{$\tick$\hss}\hrulefill \hbox to 7.5pt{$\tick$\hss}$}

\def\dtick{\leaders\hrule height .34pt depth 0.5ex \hskip 0.5pt}
\def\downboxfill{$\moth \setbox\zo\hbox{\dtick}%
  \hskip 2pt\hbox to 0pt{$\dtick$\hss}\hrulefill \hbox to 2pt{$\dtick$\hss}$}

\def\bec{\begin{center}}
\def\ec{\end{center}}

\def\nn{\nonumber}

\def\tr{{\rm tr}}
\def\Tr{{\rm Tr}}

\def\be{\begin{equation}}
\def\ee{\end{equation}}
\def\bea{\begin{eqnarray}}
\def\eea{\end{eqnarray}}
\def\ba{\begin{array}}
\def\ea{\end{array}}

\begin{document}

\begin{titlepage}
	\rightline{}
	%\rightline\today 
	\rightline{April  2023}
	\rightline{HU-EP-23/07-RTG} 
	\rightline{MIT-CTP/5553} 
	\begin{center}
		\vskip 1.1cm
		 
		 {\Large 
		 \bf{2D Black Holes, 
		 Bianchi I Cosmologies, and $\alpha'$		  }}
		
		\vskip 1.5cm

{\large\bf {Tomas Codina$^\dag$, Olaf Hohm$^\dag$ and Barton Zwiebach$^*$}}
		\vskip 1cm
		
		$^\dag$ {\it   Institute for Physics, Humboldt University Berlin,\\
			Zum Gro\ss en Windkanal 6, D-12489 Berlin, Germany}\\
		
		\vskip .3cm
		
		$^*$ {\it   Center for Theoretical Physics, \\
		Massachusetts Institute of Technology, \\
		Cambridge MA 02139, USA}\\
		\vskip .1cm
		
		\vskip .4cm

		tomas.codina@physik.hu-berlin.de, ohohm@physik.hu-berlin.de, zwiebach@mit.edu

		\vskip .3cm
		\end{center}

	\bigskip\bigskip
	\begin{center} 
		\textbf{Abstract}

	\end{center} 
	\begin{quote}  
		
We report  two surprising results on  
           $\alpha'$ corrections in string theory
           restricted to massless fields. 
          First, for 
          critical dimension  
          Bianchi type I  cosmologies 
          with $q$ scale factors only $q-1$ of them 
          have  non-trivial 
          $\alpha'$ corrections. In particular,  for 
          FRW  backgrounds
          all $\alpha'$ corrections are 
           trivial. 
          Second,  in non-critical dimensions, 
          all terms in the spacetime action
          other than the cosmological term 
          are field redefinition equivalent to terms with 
          arbitrarily many derivatives, with the latter generally of the same order.           
          Assuming 
          an $\alpha'$ expansion 
          with coefficients that fall off sufficiently fast, we consider field
          redefinitions consistent with this fall-off and classify the higher derivative 
          terms 
          for two-dimensional  
          string theory with one timelike isometry.
          This most general duality-invariant theory permits black-hole solutions,
          and we provide perturbative and non-perturbative tools to explore them.

	\end{quote} 
	\vfill
	\setcounter{footnote}{0}
\end{titlepage}

%\tableofcontents

%\hfill \today

\newpage

\tableofcontents

\section{Introduction}

String theory features a fundamental length scale, which is the square root of the inverse string tension $\alpha'$ and 
expected to be roughly $\sqrt{\alpha'}\sim 10^{-32}\,{\rm cm}$.   
Some of the mysteries of string theory revolve around the precise ramifications of this possibly minimal 
length scale (see, for example,~\cite{Witten:1996mlx}). 
In this paper we report on two surprising results (at least surprising to the authors) regarding 
the higher-derivative corrections  of Einstein gravity that come from string theory and  that  are governed 
by $\alpha'$. One pertains to 
cosmologies in critical string theory \cite{Brandenberger:1988aj,Veneziano:1991ek,Gasperini:1996fu} 
and  one to black holes in two-dimensional 
string theory~\cite{Witten:1991yr, Mandal:1991tz,Giveon:1991sy}.

In order to determine the higher-derivative corrections,   
the first step is to find, up to a given order in $\alpha'$, a basis of higher-derivative invariants, 
i.e., to  classify the independent higher-derivative terms up to field redefinitions. 
Apart from assumptions of locality and invertibility, the field redefinitions to be investigated are those that respect the 
symmetries one assumes the higher-derivative terms to have. 
The effective field theory of strings can be written 
consistent with manifest 
diffeomorphism invariance, and so  one considers field 
redefinitions that preserve the tensor character of fields. For instance, the Riemann-squared term of bosonic or heterotic string theory 
in critical dimensions cannot be changed by  
field redefinitions, and hence its coefficient 
has an invariant meaning.\footnote{It should be emphasized, however, that field redefinitions that do \textit{not} preserve the tensor character 
are perfectly legal and may, in fact, be necessary when additional symmetry principles are imposed, 
as for instance in double field theory with $\alpha'$ corrections \cite{Siegel:1993th,Hull:2009mi,Hohm:2010pp,Hohm:2013jaa,Hohm:2014xsa}.}

In this paper we consider string theories   in critical and non-critical dimensions. 
For simplicity, we focus on classical bosonic strings  
 and we will examine cosmological backgrounds in the critical dimension as well as the familiar $D=2$ (two spacetime dimensional) black hole.\footnote{In string theory usage, the black hole background is generally considered a critical string theory, as one is working directly with a theory
of matter central charge $26$.  The name non-critical strings is reserved for non-conformal field theories coupled to two-dimensional gravity, in which case the Liouville mode of the metric helps restore conformal invariance.} 
Our work will consider low-energy limits of these string backgrounds.  
 We will discuss situations where the total number of spacetime dimensions $D$ is written as $D=d+1$,  with $d$ abelian isometries 
and with fields depending only on the one remaining coordinate, which may
be timelike or spacelike.  In this situation,  the effective field theory of 
classical strings   
possesses  a global $O(d,d,\mathbb{R})$ duality invariance 
\cite{Meissner:1991zj,Meissner:1991ge,Sen:1991zi}. Therefore,  
it should be possible to write the higher-derivative terms in 
a manifestly $O(d,d,\mathbb{R})$ invariant form, and the field redefinitions should respect this structure.  Interestingly,  
as implied by the seminal work of Meissner \cite{Meissner:1996sa}, 
manifest $O(d,d,\mathbb{R})$ invariance is in conflict 
with manifest diffeomorphism invariance in $D$ dimensions, a fact that reveals itself 
 forcefully  in the general setting of 
double field theory \cite{Hohm:2013jaa,Hohm:2014xsa}. 
Perhaps somewhat surprisingly, 
in the context of dimensional reduction to one dimension a complete classification of all duality invariant higher-derivative terms can be obtained and  only first-order derivatives are needed \cite{Hohm:2019jgu,Hohm:2019ccp}, see also \cite{Codina:2020kvj,Codina:2021cxh}.

Any classification of higher-derivative terms depends on the space of backgrounds or fields to be included.  
The above mentioned `cosmological' classification with fields depending only on time includes 
the $g_{00}$ component of the metric, the purely spatial components of the metric 
and $B$-field, and the dilaton.  Restricting this space of backgrounds further, there will generally be a more refined   
classification. For instance, we may assume the consistent truncation 
where  the $B$-field vanishes and the spatial metric is diagonal, 
with $d$ `scale factors' on the diagonal that may or may not be equal. For this smaller space of backgrounds there are fewer higher-derivative 
terms that one can write, but also fewer field redefinitions, so that the classification problem has to be reconsidered.   More restrictive backgrounds provide computational
advantages given the smaller configuration space, but also provide less options to consider deformations of solutions and to explore their stability.   
As one of the two main technical results of this paper we show that for one of the scale factors (that can be picked arbitrarily) 
all higher-derivative terms can be removed by field redefinitions. In particular, specializing further to the case that all scale 
factors are equal, corresponding to Friedmann-Robertson-Walker (FRW) backgrounds, 
it follows that 
 \textit{all} higher-derivative terms are removable by field redefinitions. 
The non-perturbative cosmological FRW backgrounds with a single scale factor explored in 
\cite{Hohm:2019jgu,Hohm:2019ccp} were obtained in the context of $\alpha'$ corrections
that cannot be removed for general time-dependent backgrounds.  In that context, general time-dependent perturbations of the solution can be consistently analyzed. 
Moreover, it should also be emphasized that the removal of higher-derivative terms is strictly perturbative, 
so that there may be non-perturbative solutions that are not accessible for classifications of very restrictive backgrounds. 
More generally, perturbations or fluctuations away from a background  may not preserve any conditions, as 
in cosmological perturbation theory, where the fluctuations depend on all coordinates. 
In order to use  dualities one then requires a genuine  double field theory \cite{Hohm:2022pfi}.

As the second main result of this paper we revisit the subject of  higher-derivative modifications of string theory in non-critical dimension, 
with a particular focus on the black hole solution in two 
dimensions~\cite{Witten:1991yr, Mandal:1991tz,Giveon:1991sy}. In fact, one of the continuing appeals of this black hole solution is that it is based on an exact CFT, hence 
giving rise to an exact string background.  
Since the access to this exact background is quite limited,   
 $\alpha'$ corrections of the two-derivative solution were considered in~\cite{Dijkgraaf:1991ba}, and analyzed in some detail by Tseytlin in~\cite{Tseytlin:1991ht,Tseytlin:1993df} who asserted that a well-defined $\alpha'$ expansion in noncritical dimensions
does not exist.  Moreover, while string loop corrections can be made 
arbitrarily small, the  $\alpha'$ corrections cannot \cite{Karczmarek:2004bw}, hence making it particularly urgent 
to get a handle on these corrections.

We point out some surprising problems with the interpretation of higher-derivative terms as perturbative corrections to 
the leading order action or solution. 
To explain this let us recall the spacetime action for the metric and dilaton
of string theory in $D$ dimensions:  
\be\label{noncriticalspacetimeaction}  
I[g,\phi]  = \tfrac{1}{2\kappa_0^2} \int d^Dx \sqrt{-g} e^{-2\phi}\Bigl( \, - \tfrac{2(D-26)}{3\alpha'}  + R  +  4g^{\mu\nu}\partial_{\mu}\phi \partial_{\nu}\phi  
   +\sum_{n=1}^{\infty}  (\alpha')^n  {F} ^{(n)}(g,\phi)   
\Bigr) \,,   
\ee
where the $F^{(n)}$ denote  possible terms of order $2n+2$ in derivatives, and we will sometimes refer to the term proportional 
to $\frac{1}{\alpha'}$ as the cosmological term. We will show that for $D\neq 26$ 
{\em any}  
  term in the action other than the cosmological  term 
can be traded, by means of field redefinitions involving derivatives, 
for a term with higher derivatives. 
This includes 
the two-derivative Einstein-Hilbert term that could be traded for a term with four derivatives.  
Iterating the redefinitions, the Einstein-Hilbert term could be traded for
terms with, say, 
42 derivatives. 
Alternatively, field redefinitions allow us to rewrite the theory with the original
cosmological term and interactions, as a theory with the cosmological term
and interactions having $2n$ or more derivatives, for
any $n \geq 2$.   
In contrast to string theory in the critical dimension, 
there are \textit{no}  invariant terms 
apart from the cosmological term.

This result is puzzling, because adopting the usual perturbative mindset one would view terms with large numbers of derivatives as sub-leading compared 
to a term with two derivatives, and hence one would feel free to  drop them. This is indeed the standard procedure of bringing  higher-derivative terms 
to a minimal form, but using this procedure literally 
for string theory away from the critical 
dimension, one would conclude that only the cosmological term is non-trivial. 
 What is really happening is that in such string backgrounds 
generic  higher-derivative term are not actually 
 sub-leading relative to terms with lesser number of derivatives. 
 Thus, while the field redefinitions to be discussed 
 are perfectly legal, it is the {\em second step} of dropping induced terms with more 
 derivatives that is generally illegal. 
 
 To elaborate  
  this point let us recall 
  that  while informally one refers to the higher-derivative corrections of string theory as 
 `\,$\alpha'$ corrections\,', $\alpha'$ itself, being dimensionful,  is not a small expansion parameter. 
In terms of the fundamental length scale $\sqrt{\alpha'}$
of string theory, $\alpha'$ is just 
one. 
In fact, $\alpha'$ can be eliminated  from the action  (\ref{noncriticalspacetimeaction}) by 
defining the dimensionless derivative operator 
 \be
  \bar{\partial}_{\mu} := \sqrt{\alpha'} \frac{\partial}{\partial x^{\mu}} \;, 
 \ee 
and rescaling the action by $\alpha'$:  
 \be
  \bar{I} := \alpha' I 
  = \tfrac{1}{2\kappa_0^2} \int d^Dx \sqrt{-g} e^{-2\phi}\Bigl( \, - \tfrac{2(D-26)}{3}
  + \bar{R} + 4g^{\mu\nu} \bar{\partial}_{\mu}\phi \bar{\partial}_{\nu}\phi 
  +\sum_{n=1}^{\infty}   \bar{F} ^{(n)}(g,\phi)  \Bigr) \;,   
 \ee  
where the  bar 
over $R$ or $F$ indicates that all derivatives $\partial_{\mu}$ have been replaced by $\bar{\partial}_{\mu}$. 
In this formulation there is no $\alpha'$ left, and there is no expansion in $\alpha'$. Rather, 
one should think of the higher-derivative corrections as an expansion 
in terms of small derivatives of the fields. While this can make   sense in critical-dimension string theory, 
in non-critical dimension string theory generic solutions feature fields whose 
dimensionless derivatives are of order $\bar{\partial}\sim {\cal O}(1)$, 
so that all higher-derivative terms can have significant effects.

Indeed,  the 2D black-hole solution is obtained by balancing the effects of the
order $1/\alpha'$ cosmological term and the two derivative terms,
the result being a configuration where dimensionless derivatives are of order one. 
This 
sheds doubt on attempts to find a more accurate black hole solution by means of higher-derivative 
corrections.  Nevertheless,  we show that if the higher derivative terms are
suppressed in a particular way, some simplifications of the effective action 
are valid.  
To see this, suppose 
an oracle gives us an action of the form (\ref{noncriticalspacetimeaction}) with infinitely many  higher-derivative terms. 
A priori, general higher-derivative terms are   
all  of the same order.
Let us suppose, however,  that the higher-derivative 
terms  come with numerical coefficients 
that fall off in such a way that terms with four or more  
 derivatives are  sub-leading compared to terms with less 
derivatives.  
 Since two-derivative terms come  
with order one coefficients, this could happen if the terms
of order $(\alpha')^n$, with $n\geq 1$, come with coefficients 
of order $\epsilon^n$ with $\epsilon < 1$.   
In this situation we can ask and answer the following question:  
What are the most general field redefinitions that 
preserve this pattern, and what are the most general higher-derivative corrections modulo these restricted field redefinitions?   
We will show that these additional requirements eliminate those field redefinitions that allow one to remove arbitrary terms,  and we will arrive at a minimal non-trivial set of higher-derivative terms that resembles the cosmological classification 
for critical string theory.   Given the current knowledge of derivative  
corrections, we cannot know if such a classification applies to the 
$2D$ black hole.  
The general action in this setup, where fields are time independent, 
 is obtained in section~\ref{sec_classification} and takes the form
\begin{equation}\label{classificationintro}
I = \int dx \, n\,  e^{-\Phi} \biggl[ Q^2   
+ (\D\Phi)^2 - M^2 + \sum_{i\geq 1} \frac{\epsilon_i}{Q^{2i}}  M^{2i + 2} \biggr]\,. 
\end{equation}
Here the metric is $ds^2 = - m^2 (x) dt^2  + n^2 (x) dx^2$,  we defined  
$M= \tfrac{1}{n} \partial_x \ln m$, $Q^2 = 16/\alpha'$, and $\Phi$ is the duality-invariant dilaton.  The first three terms in brackets define the action 
up to two derivatives;
the last term, with arbitrary coefficients, 
represents the possible inequivalent higher-derivative terms.  General field redefinitions of the `lapse' function $n(x)$ are not allowed for the classification --- those are redefinitions that can remove any higher-derivative term.  A linear combination of the lapse and dilaton can be redefined, and
so can the metric component~$m$.   
The work in~\cite{Ying:2022xaj} assumed an expansion
analogous to that above and discussed possible solutions of the resulting equations,
aiming to resolve the black hole singularity (the case when the metric depends on a single spatial coordinate is found in~\cite{Wang:2019mwi}).

This paper is organized as follows. In section~\ref{sectiontwooo} we point out and discuss the subtleties 
that arise for field redefinitions of the string effective action in non-critical dimensions, i.e., in presence 
of the cosmological term.   
We do this both in general dimensions and in dimensional reduction to one spatial dimension where, helped by duality symmetry  
and the simplicity of the theory, we classify higher-derivative  terms up to the above mentioned  class of field redefinitions.
In section~\ref{blddcljkdfoiw} we revisit the two-dimensional black hole solution and discuss possible perturbative and non-perturbative 
$\alpha'$ modifications.  We also discuss the proposal of 
Dijkgraaf, Verlinde, and Verlinde~\cite{Dijkgraaf:1991ba} for a possibly exact background. In section~\ref{bianchioneone} we revisit string cosmologies of type Bianchi I in critical dimensions and classify all 
higher-derivative corrections.   We offer some concluding remarks in section~\ref{concleoridk}.

\section{String theory in $D\not= 26$ and in $D=2$}  \label{sectiontwooo}

In this section we begin by showing how in the $D\not= 26$ 
string effective action for massless fields, 
having a cosmological term of order $1/\alpha'$, 
any interaction term can be removed
by a field redefinition of the metric.  We then turn to the important case
when the spacetime dimension is two ($D=2$) and examine the theory
with the assumption that fields do not depend on time --- they only depend
on the spatial coordinate $x$.  The field variables are the `lapse' function
$n(x)$ whose square multiplies $dx^2$ in the metric, a metric component $m(x)$,
whose square multiplies
$dt^2$, and a duality invariant dilaton $\Phi$.  
We use a simplified `dilaton-lapse'
model to discuss possible field redefinitions, noting that they fall into two classes,
one in which the cosmological term varies, and one in which it does not. 
We find that solutions, like the black hole  
or those of the dilaton-lapse model, do not lend themselves to an $\alpha'$ expansion.
Instead we discuss a possible suppression of derivative corrections that could
allow for a consistent set of field redefinitions --- those in the second class above.  
We use this to finally give a classification of higher-derivative interactions
for the $D=2$ backgrounds that have no time dependence.

\subsection{Field redefinitions for $D\not= 26$} \label{sec_ID}  

Let us reconsider the effective spacetime action $I_D$ for strings 
in $D$ dimensions (\ref{noncriticalspacetimeaction}) including its higher derivative corrections: 
 \be
 \label{full-action}
 \begin{split}
  I_D[g,\phi]   
   &= \int d^Dx \sqrt{-g} e^{-2\phi}\biggl( -\Lambda + R+ 4g^{\mu\nu}\partial_{\mu}\phi \partial_{\nu}\phi  
  +\sum_{n=1}^{\infty} (\alpha')^n  F^{(n)}(g,\phi)  \biggr)\,, 
 \end{split}
 \ee 
where we have set the $B$-field to zero, we have introduced the constant 
$\Lambda$ defined to be
\be
\Lambda =   \frac{2(D-26)}{3\alpha'}\,, 
\ee
and the  $F^{(n)}$ are arbitrary functions of $g$ and $\phi$ 
of order $2n+2$ in derivatives.   Note that the cosmological term 
is of order $1/\alpha'$, the two-derivative terms are of zeroth order in
$\alpha'$, and the higher derivative terms begin at order $\alpha'$. 

Consider now the {\em exact} field redefinition of the metric
 \be
  g_{\mu\nu} \rightarrow   
   g_{\mu\nu}  + \Delta g_{\mu\nu} \;, 
 \ee
where we will take $\Delta g_{\mu\nu}$ to be a local function given by a derivative expansion.\footnote{As a field redefinition one must view this replacement
as setting $g_{\mu\nu} =  g'_{\mu\nu} + \Delta g_{\mu\nu} ( g')$ so that the action becomes
$ I (g) = I (g' +\Delta g(g')) \equiv I'(g') $.  One can solve for $g'$ in terms of $g$ by
inverting the expression for $g$ in terms of $g'$. }  
 This implies 
 \be\label{redefgs}
 \begin{split}
  g^{\mu\nu}  &\to \
   g^{\mu\nu} - \Delta g^{\mu\nu} + {\cal O} ((\Delta g)^2) 
   \,,  \\ 
  \sqrt{-g}  & \to \ \sqrt{-g} \Bigl(1+\tfrac{1}{2} g^{\mu\nu} \Delta g_{\mu\nu}
  + {\cal O} ((\Delta g)^2) \bigr)  
  \\
  R  &\to \  R +\Delta g^{\mu\nu} \Bigl( R_{\mu\nu}-\tfrac{1}{2} g_{\mu\nu} R\Bigr) + g^{\mu\nu} (\nabla_{\rho} \Delta \Gamma^{\rho}_{\mu\nu}
  -\nabla_{\mu} \Delta \Gamma_{\rho\nu}^{\rho}\big) +\ {\cal O} ((\Delta g)^2)  \,.
 \end{split} 
 \ee
Indices on $\Delta g_{\mu\nu}$ are raised with the unperturbed $g^{\mu\nu}$. 
We take $\Delta g_{\mu\nu}$ to be given by a derivative expansion: 
 \be
  \Delta g_{\mu\nu} = \Delta^{(1)}  g_{\mu\nu} + \Delta^{(2)}  g_{\mu\nu} + \cdots \;, 
 \ee
where $\Delta^{(n)}  g_{\mu\nu}$ is  of order $2n+2$ in derivatives. The first term
in the redefinition above has four derivatives.

The redefined action $I'_D$ is given by (\ref{full-action}) with $g$ replaced by $g + \Delta g$: 
 \be\label{redefinedI} 
 \begin{split}  
  I'_D := I_D[g+ \Delta g, \phi]  = \int  d^Dx &  \sqrt{-g} e^{-2\phi}\Bigl( -\Lambda + R+ 4g^{\mu\nu}\partial_{\mu}\phi \partial_{\nu}\phi \Bigr) \\
    &+ \tfrac{1}{2} \int d^Dx \sqrt{-g} e^{-2\phi} g^{\mu\nu} \Delta^{(1)}  g_{\mu\nu}   (-\Lambda) 
    \\
  &+ \alpha'  \int d^Dx \sqrt{-g} e^{-2\phi} \, F^{(1)}(g,\phi)  +\cdots 
   \;, 
 \end{split} 
 \ee
where we used (\ref{redefgs}) and where  the ellipsis denote terms with \textit{more than four  derivatives}. 
This follows from $\Delta^{(1)}  g_{\mu\nu}$ being already of fourth order in derivatives, so that 
in particular new terms induced from the two-derivative action  are already of order six  in derivatives.
We can cancel the four-derivative $F^{(1)}$ term by choosing  
 \be
  \Delta^{(1)}  g_{\mu\nu} =   {2\over D} \, {\alpha'\over \Lambda}  \, g_{\mu\nu} F^{(1)}(g,\phi) \,.
 \ee
Since $\Lambda$ is of order $1/\alpha'$, the above right hand side is of order
$\alpha'^2$.  The six-derivative terms encoded 
in $F^{(2)}$ receive further contributions from the field redefinition, and we denote the totality of all such terms by $\tilde{F}^{(2)}$. 
The redefined action then reads  
\be
\label{afterfirstredef}
I'_D = \int d^Dx \sqrt{-g} e^{-2\phi}\left( - \Lambda 
+ R+ 4g^{\mu\nu}\partial_{\mu}\phi \partial_{\nu}\phi  
  + (\alpha')^2 \tilde{F}^{(2)}(g,\phi) +\cdots \right) \;, 
\ee
where the ellipsis denotes all terms with more than six derivatives.    Of course,
we could have instead cancelled the Einstein-Hilbert term $R$ by including
a two-derivative term $\Delta^{(0)} g$ in the $\Delta g$ expansion, and setting
$\Delta^{(0)} g_{\mu\nu} = {2\over D} \, {1\over \Lambda}  \, g_{\mu\nu} R$.
In that case the action would have the cosmological term followed by terms
with four derivatives.  

The procedure above can be iterated.  Looking at the action~(\ref{afterfirstredef}) 
we can just repeat the procedure by setting 
 \be
    \Delta^{(2)}  g_{\mu\nu} =  {2\over D} {\alpha'^2 \over \Lambda}  \,  g_{\mu\nu} \, \tilde{F}^{(2)}(g,\phi)\;,  
 \ee
so as to cancel the terms $\tilde{F}^{(2)}$ with six derivatives. 
Thus, all higher-derivative corrections can be moved to arbitrary high order in $\alpha'$.

\subsection{Dimensional reduction} \label{dimreduvmbb}

We want to analyze the duality properties and $\alpha'$ corrections  of the black hole solution in 2D string theory \cite{Witten:1991yr,Mandal:1991tz}.
The two-derivative  action for string theory in $D$ dimensions, taken with vanishing $B$-field,
 reads 
 \be
 \label{orig-action}
  I_D = \int d^Dx \sqrt{-g} e^{-2\phi}\left( \,  - \tfrac{2(D-26)}{3\alpha'} 
  + R+ 4g^{\mu\nu}\partial_{\mu}\phi \partial_{\nu}\phi \right)\,.
 \ee 
 To focus on the string theory black hole we set $D=2$,  
 with coordinates $x^{\mu}= (x^0, x^1) = (t,x)$ and fields that do not depend on time $t$.  We thus
 make the ansatz
 \be
  g_{\mu\nu} = \begin{pmatrix} -m^2(x) & 0 \\ 0 & n^2(x) \end{pmatrix} \;, \qquad \phi= \phi(x)\;. 
 \ee 
It is worthwhile to compare with the cosmological case, where all fields depend on time
and are independent of $d$ internal spatial coordinates.  In this case one has a global $O(d,d)$ duality symmetry.  Here, spacetime is two dimensional, and the fields do not
depend on time.  Time is then the one `internal' coordinate and the duality group is just $O(1,1)$.  In the cosmological setting, the component of the metric in the time-time
direction is the lapse function.  Here, the component $n(x)$ of the metric in the space-space direction is the analog of the cosmological lapse function.
The resulting action will be $x$-reparameterization invariant.  With 
reparameterizations $x \to x - \lambda (x)$ we have that scalars $A$ transform as $\delta_\lambda A = \lambda \partial_x A$.  The field $n(x)$ transforms as a density:
$\delta_\lambda n = \partial_x (\lambda n)$.    
Since only $x$ derivatives exist, we do not need partial derivatives, and ordinary $x$ derivatives will
be denoted with primes ${}^{ \prime}  \equiv \frac{d}{dx}$.  
When multiplied with $n^{-1}$, $x$-derivatives then  give the  covariant derivative
 \be\label{covxder}
  D\equiv  \frac{1}{n} \frac{d}{dx}\;. 
 \ee
 If $A$ is a scalar, then $DA$ is also a scalar. 
 The group of dualities here is $O(1,1)$.  In the component connected
 to the identity, group elements $h$ are of the form 
 \be
 h = \begin{pmatrix} e^\alpha & 0 \\ 0 & e^{-\alpha} \end{pmatrix} \,, \ \ \  h^t \eta h = \eta\,, \ \ \hbox{with} \ \   \eta = \begin{pmatrix} 0 & 1 \\ 1 & 0\end{pmatrix}  = \eta^t = \eta^{-1} \,.  
 \ee
Here $\alpha$ is any real number, and $\eta$ is the $O(1,1)$ metric.  
In the component disconnected to the identity, 
we have  
\be
h = \begin{pmatrix} 0 & e^\alpha  \\  e^{-\alpha} & 0 \end{pmatrix} \in  O(1,1) \,. 
\ee
Note that for $\alpha = 0$ we have $h = \eta$, as a group element. 

 In this setup, the generalized metric ${\cal S}$
is a two-by-two matrix that involves the internal metric components, that is, the
component $m^2$ introduced above, and the 
internal $B$-field,  
which vanishes in
one dimension.  We thus get  
\be
  {\cal S} = \begin{pmatrix} 0 & m^2 \\ m^{-2} & 0 \end{pmatrix} \ \in \ O(1,1)\,,
 \ee 
 where we noted that ${\cal S}$ is an element of the group $O(1,1)$, in fact, an element belonging in  the component disconnected to the identity, as it is clear from its determinant being equal to minus one.  Under duality one has 
 \be
 \label{s-fduality}
 {\cal S}\ \to \  h\,  {\cal S} \, h^{-1}\,.
 \ee
 For the duality transformations connected to the identity, 
  the field $m$ is scaled by a constant: 
  $m\to m e^\alpha$.  
  The field $n(x)$ is
  duality invariant, and we have $\phi\to \phi + \tfrac{\alpha}{2}$.  
   As a result,
  we  have  the duality-invariant dilaton $\Phi$ given as
 \be
 \label{dil-dil}
  e^{-\Phi(x) } \equiv m(x) e^{-2\phi(x)}\,,  
 \ee
making the duality invariance of the measure clear:
 \be
  \sqrt{-g}e^{-2\phi} = n m e^{-2\phi} =  n e^{-\Phi}\;.
 \ee 
For the disconnected dualities we take $h = \eta$ and then have  
\be
\label{discretedu} 
{\cal S} \to  \eta {\cal S} \eta  =  \begin{pmatrix} 0 & m^{-2} \\ m^{2} & 0 \end{pmatrix}\,, 
 \ \quad  \hbox{or} \ \quad   m \to  {1\over m} \,,  
\ee
while invariance of (\ref{dil-dil}) yields $\phi\rightarrow \phi - \ln |m|$. 
This is the familiar discrete $\mathbb{Z}_2$  duality.

The effective action can be written ignoring time, as no quantity is time dependent.
The resulting  one-dimensional action, setting $D=2$ in~(\ref{orig-action})
and on account of the above comments, takes the form
 \be  
  I = \int dx \,n \, e^{-\Phi}\left(\,   Q^2 + R + 4 n^{-2} (\phi')^2 \right) \,,
 \ee
 where we defined  
 \be
 Q^2 \equiv  {16\over \alpha'} \,. 
 \ee
 To get a useful expression in terms of the $m$ and $\Phi$ fields, we need to work out the Ricci scalar for the above metric ansatz. The non-vanishing Christoffel symbols are given by 
 \be\label{nonzeroChristoffel}
  \Gamma_{00}^{1} = \frac{1}{n^2} m  m' \;, \qquad \Gamma_{10}^{0} = m^{-1}{m'}\;, \qquad \Gamma_{11}^{1} = n^{-1} n'\;. 
 \ee 
The Ricci tensor 
 $R_{\nu\sigma} = \partial_{\mu}\Gamma^{\mu}_{\nu\sigma} -\partial_{\nu}\Gamma_{\mu\sigma}^{\mu} +\Gamma_{\mu\lambda}^{\mu}
  \Gamma_{\nu\sigma}^{\lambda} -\Gamma_{\nu\lambda}^{\mu} \Gamma_{\mu\sigma}^{\lambda}$
then yields the components 
 \be
    R_{00} = \frac{1}{n^2} (m m'' -n^{-1}n' m m')\;,\ \ \ 
    R_{11} = -m^{-1}m'' +m^{-1}m' n^{-1}n'\; .
 \ee 
The Ricci scalar is therefore
 \be\label{CURVATURE}
  R= -\frac{2m''}{mn^2} + \frac{2}{mn^3} n' m'  = - {2\over mn} \Bigl( {m'\over n} \Bigr)' 
  = - {1\over mn}  \Bigl( {(m^2)' \over mn} \Bigr)'  \;. 
 \ee
Inserting this into the action, using~(\ref{dil-dil}) to pass from $\phi$ to $\Phi$ and $m$,  and integrating by parts in order to have only first-order derivatives gives 
 \be
  I = \int dx \,\frac{1}{n}\, e^{-\Phi}\left(\, n^2 Q^2+\, (\Phi')^2 - (m^{-1}m')^2
  \right)\;.  
 \ee
The action can also be written in terms of the generalized metric ${\cal S}$
 \be
  I = \int dx \,\frac{1}{n}\, e^{-\Phi}\left( \,   n^2 Q^2  
  + (\Phi')^2  +\tfrac{1}{8} {\rm tr}\big({\cal S}'\big)^2  \right)\;.
 \ee
$O(1,1)$ global duality invariance is manifest because $n$ and $\Phi$ are invariant and
${\cal S}$ transforms as indicated in~(\ref{s-fduality}).   Finally, defining  
\be
   M\equiv m^{-1} D m  =  {m'\over mn} \;,  \ee
we can also write the action as   
\be
\label{lforma}
I = \int dx \, n\,  e^{-\Phi} \bigl(  Q^2 +  (D\Phi)^2 - M^2 \bigr)\,. 
\ee

The equations of motion follow from the general variation 
\be
\label{gen-var-action}
\delta I = \int dx \, n \, e^{-\Phi} \left[ \,  {\delta n\over n}  \, E_n +  2 \frac{\delta m}{m}\, E_m  + \delta \Phi \,E_{\Phi} \, \right]  \,, 
\ee  
for which we find 
 \be\label{simplifiedequations}
  \begin{split}  
   E_n &= -(D\Phi)^2 + M^2 + Q^2 = 0\;, \\ 
     E_{m} & = DM-(D\Phi) M    = 0\;, \\
   E_{\Phi} &= -2D^2\Phi  + (D\Phi)^2 + M^2 -  Q^2= 0  \;. 
   \end{split}
 \ee 
As a consequence of coordinate-reparameterization invariance, these equations are not all independent but  satisfy the Bianchi identity 
\begin{equation}\label{BI}
D \Phi \left(E_n + E_{\Phi}\right) + 2 M E_m - D E_n = 0\,.
\end{equation}
We note that 
the second equation in (\ref{simplifiedequations}) implies 
 \be
   (e^{-\Phi}  M )' = e^{-\Phi} (M'-\Phi' M) 
   = 0\;,  
 \ee
i.e., that $e^{-\Phi} M$ is constant and does not depend on $x$. This is a manifestation of Noether's 
theorem for  $O(1,1)$ invariance.

\subsection{Two classes of field redefinitions}

In this section we will explore field redefinitions of  
 a dilaton-lapse  model with a cosmological term. 
In this model, obtained by setting $M=0$ in the action~\eqref{lforma}, 
one can easily find solutions of the equations of motion, even after
including a large class of higher derivative terms.  This allows us
to see explicitly the effect of field redefinitions.   Moreover, the results obtained
here apply almost without change to the case of the $D=2$ black hole.

We will see clearly in this dilaton-lapse model that solutions
of the lowest order nontrivial equations imply that a conventional
derivative expansion is problematic: there is no
obvious suppression of the higher derivative terms. All terms
seem equally important and field redefinitions used to classify 
interactions do not operate as usual. 

Our analysis below will assume theories in which there is 
an {\em in-built} suppression of the higher derivative terms, a suppression
due to constants that multiply the interactions and become
smaller as the number of derivatives increase. 

While we cannot justify such an assumption in the case of the 
2D black hole, the assumption does seem to hold for certain field
theories arising from string theory, as is the case of tachyon dynamics,
where, as discussed in~\cite{Erbin:2021hkf}, the Lagrangian $L$ takes the
form
\be
L =  \tfrac{1}{2} \phi ( \partial^2 +1 ) \phi  + \tfrac{1}{3} \bigl( e^{\xi^2 \partial^2} \phi \bigr)^3 \,. 
\ee
This Lagrangian, written in terms of unit-free fields and derivatives, depends on a
constant $\xi$, a non-locality parameter.  Roughly, a term with $2k$ derivatives comes
with a factor $\xi^{2k}$.  If $\xi < 1$ there is some suppression, and for $\xi \ll 1$ a 
strong suppression. 

Returning to our dilaton-lapse model, even assuming an in-built
suppression,  complications
arise due to the cosmological term.  We will argue that there are two classes of field redefinitions:

\begin{enumerate}

\item[1.]  Separate field redefinitions of the lapse or dilaton function, which generate
new interactions via the variation of the cosmological term as well
as the variation of other terms.

\item[2.]  Simultaneous field redefinitions of the lapse and dilaton for
which no terms arise from the variation of the cosmological term.

\end{enumerate} 

We will argue that redefinitions of type 1 do not preserve the structure
of in-built suppression:  higher derivative terms induced by the redefinitions
are not suppressed appropriately and thus cannot be neglected.
Redefinitions of type 2, however, respect the structure of in-built 
suppression and thus can be used in the conventional setting of
effective field theory to classify interactions.    We will consider 
the issue first on a simplified model and then we will generalize.

\noindent
{\bf  Dilaton-lapse model and redefinitions.} 
Consider, therefore,  the 2-derivative theory obtained by
setting $M=0$ in the action~\eqref{lforma}:
\begin{equation}\label{ID2}
I^{(0)} = \int dx \, n\,  e^{-\Phi} \left[ Q^2 + (D\Phi)^2 \right]\,.
\end{equation}
Its equations of motion are given by
\begin{equation}\label{EOMD2}
\begin{aligned}
E^{(0)}_{n} &\equiv Q^2 - (\D \Phi)^2 = 0\,,\\
E^{(0)}_{\Phi} &\equiv - Q^2 + (\D \Phi)^2 - 2 \D^2 \Phi = 0\,.
\end{aligned}
\end{equation}
These are solved by the linear dilaton background
\begin{equation}\label{Phi0}
n^{(0)}=1, \quad \Phi^{(0)} = Q x \,.
\end{equation}
The simplest higher-derivative extension to \eqref{ID2} is given by
\begin{equation}\label{ID4}
I^{(1)} = \int dx \, n\,  e^{-\Phi} \left[ Q^2 + (\D\Phi)^2  + c \alpha' (\D \Phi)^4\right]\,.
\end{equation}
Note that for the solution $\Phi^{(0)} = Qx$, all three terms in the action are
of the same order $Q^2$ (since $\alpha' \sim 1/Q^2$).  
The lapse equation following from the above action is given by
\begin{equation}\label{EOMDphi4}
\begin{aligned}
E_n &\equiv Q^2 - (\D \Phi)^2 - 3 c \alpha' (\D \Phi)^4 = 0\,,
\end{aligned}
\end{equation}
and the dilaton equation is automatically satisfied when the lapse equation
holds due to the Bianchi identity 
(see \eqref{BI} specialized to   
$E_m=0$). The above equation admits a unique real solution of the form
\begin{equation}
n=1, \quad \Phi = \omega x, \quad \omega^2 \equiv \frac{\sqrt{1 + 12 c \alpha' Q^2} - 1}{6 c \alpha'}\,.
\end{equation}
This solution can be considered a small correction to \eqref{Phi0} when  
$\omega^2$ has a convergent perturbative expansion in powers of $c \alpha' Q^2$.
This happens when $|12 c \alpha' Q^2| < 1$. With $Q^2 = \frac{16}{\alpha'}$,
 the only option is a small coefficient in the action, namely $|c| < \frac{1}{192}$. 
More generally, we need 
\begin{equation}\label{epsilon}
\epsilon \equiv c \alpha' Q^2 \ll 1\,.
\end{equation}
If this condition is satisfied, 
$\omega^2$ 
can be expanded in powers of $\epsilon$ and so
\begin{equation}\label{Phi1}
\Phi = \omega x = \left(1 - \tfrac32 \epsilon + \mathcal{O}(\epsilon^2)\right)Q x = \Phi^{(0)} - \tfrac32 \epsilon Q x + \mathcal{O}(\epsilon^2)\,,
\end{equation}
where the leading term is the 2-derivative solution \eqref{Phi0} and the rest are truly small corrections to it. If \eqref{epsilon} does not hold,  the 4-derivative term in the action contributes terms comparable to the 2-derivative solution,  making a derivative expansion meaningless.  
We assume that \eqref{epsilon} holds.

For simplicity, we rewrite the theory in terms of a unit-free derivative  
$\bar \partial_x \equiv (1/Q) \partial_x$, 
so that $\hD = (1/ Q) D$. In terms of these derivatives, the action~\eqref{ID4} 
becomes 
\begin{equation}\label{IbarD4}
\bar{I}^{(1)} \equiv \frac{1}{Q^2} I^{(1)}  = \int d x \, n\,  e^{-\Phi} \left[ 1 + (\hD\Phi)^2  + \epsilon (\hD \Phi)^4\right]\,.
\end{equation}
 In this notation a variation of the lapse $n \to n+ \delta n$ gives
 to leading order
 \be
 \bar I^{(1)} \to   \bar I^{(1)} + \int dx\,  n\,  e^{-\Phi} {\delta n\over n} \bar E_n  + {\cal O} ((\delta n)^2) ,
 \ \ \ \hbox{with} \ \ \ \bar E_n =  1 - (\hD\Phi)^2 - 3 \epsilon (\hD\Phi)^4\,. 
 \ee
  We now do a lapse redefinition to remove the four derivative term.  We take
\begin{equation}\label{deltan}
\frac{\delta n}{n} = - \epsilon (\hD \Phi)^4\,.
\end{equation}
The associated field redefinition is $n = n' + \delta n (n')$ and $\Phi = \Phi'$. 
The redefined action, called $\bar I'$, and written in terms of the 
new (primed) fields is given by 
\begin{equation}\label{Iprime1}
\bar{I}' = \int d x \, n'\,  e^{-\Phi'} \left[ 1 + (\hD\Phi')^2  + \epsilon (\hD \Phi')^6 
+ \mathcal{O}(\epsilon^2)+ \mathcal{O}((\delta n)^2)\right]\,.
\end{equation}
The field redefinition eliminated the four-derivative term at the cost of introducing
a six-derivative term, {\em also} at order $\epsilon$.   Since derivatives $\bar D$ in the unperturbed theory are of order one, this new term is not parametrically smaller
than the original four-derivative term.
This shows that pure lapse transformations do not allow us 
to classify interactions in the sense of effective field theory.  They are redefinitions
of type 1, and as claimed do not respect the structure of in-built suppression. 
We would have wanted the new six-derivative interaction to appear at a higher
order in $\epsilon$. 

We now discuss type 2 transformations; those that preserve the structure
of in-built suppression and thus can be used 
to remove higher-derivative terms without inducing same-order effects. 
In this case, it will allow us to eliminate the four-derivative term consistently. 
The redefinition is of the form 
\begin{equation}\label{goodredefx}
\Phi = \Phi' + \delta \Phi (n', \Phi') \,, \quad  n= n' + \delta n (n', \Phi') \,, \quad 
\hbox{with} \ \   \delta \Phi = \frac{\delta n}{n}  \,.
\end{equation}
For such a correlated redefinition of the dilaton and the lapse, the variation
of the action to linearized order is, from~(\ref{gen-var-action}),  
\be
\label{gen-var-acion}
\delta I = \int dx \, n \, e^{-\Phi} \left[ \,  ( E_n +E_\Phi) \delta \Phi  \, \right]  \,, 
\ \  E_n +E_\Phi = -2 \bar D^2 \Phi + {\cal O} (\epsilon) \,. \ee  
An important fact is that this linear combination of equations of motion has
no constant term.  
We will choose, to order $\epsilon$, the following variation in order to remove
the four-derivative term from~\eqref{IbarD4}:  
\begin{equation}\label{goodredef}
 \frac{\delta n}{n} = \delta \Phi = \tfrac32 \epsilon (\hD \Phi')^2\,. 
\end{equation}
Under this redefinition, \eqref{IbarD4} changes to a new action, with primed fields
\begin{equation}\label{Iprime}
\begin{aligned}
\bar{I}' &= 
 \int d x \, n'\,  e^{-\Phi'} \left[ 1 + (\hD\Phi')^2  + \epsilon (\hD \Phi')^4 + \left(-2 \hD^2 \Phi'\right) \left[\tfrac32 \epsilon (\hD \Phi')^2\right] +  \mathcal{O}(\epsilon^2)\right]\\
&= \int d x \, n'\,  e^{-\Phi'} \left[ 1 + (\hD\Phi')^2  + \epsilon \left[(\hD \Phi')^4 - 3 (\hD \Phi')^2 \hD^2 \Phi'  \right] +  \mathcal{O}(\epsilon^2)\right]\\
&= \int d x \, n'\,  e^{-\Phi'} \left[ 1 + (\hD\Phi')^2  +  \mathcal{O}(\epsilon^2)\right]\,.
\end{aligned}
\end{equation}
 In the last equality we used integration by parts to see that the $\mathcal{O}(\epsilon)$ term is identically zero. This time we succeed in eliminating the 4-derivative term 
 since the only additional terms generated are truly higher-order ${\cal O} (\epsilon^2)$ effects! 

\medskip
\noindent
{\bf The general dilaton-lapse model and field redefinitions.}  The lessons of the above discussion can be refined by considering the 
general version of the dilaton model:
\begin{equation}\label{Idilaton}
I = \int dx \, n\,  e^{-\Phi} \Bigl[ Q^2 + (\D\Phi)^2  + \sum_{i\geq1} c_i \alpha'{}^i \mathcal{L}^{(2 i + 2)}(\D ;\Phi) \Bigr]\,,
\end{equation}
which includes infinitely many $\alpha'$ corrections depending only on 
covariant derivatives
of $\Phi$, with $\mathcal{L}^{(2 i + 2)}(\D;  \Phi)$ containing $2 i + 2$ derivatives. The same action can be rewritten in terms of the dimensionless derivative $\bar D$, absorbing a factor of $Q^2$: 
\begin{equation}\label{Ibardilaton}
\bar{I} = \int dx \, n\,  e^{-\Phi} \Bigl[ 1 + (\hD\Phi)^2  + \sum_{i\geq 1} \epsilon_i \mathcal{\bar L}^{(2 i + 2)} \
\bigr]\,, \ \ \  \epsilon_i \equiv c_i (\alpha' Q^2)^i\,.
\end{equation}
Again, this action has no meaningful derivative expansion unless the coefficients 
$\epsilon_i$ decay fast enough. This can be formalized by extending the condition $\epsilon \ll 1$ considered before to the condition:
\begin{equation}\label{condition}
\epsilon \equiv \epsilon_1 \ll 1\,, \quad \epsilon_i \sim (\epsilon)^i\,,  \  i \geq 1 \,, 
\end{equation}
where the symbol $\sim$ denotes proportionality up to factors of order one. 
The above condition guarantees that each term in the derivative expansion 
is parametrically smaller than the previous one. In order to explore the effect of perturbative field redefinitions, we will assume that 
this condition is satisfied.  Note that the condition above also implies that
\be
\label{multip} 
\epsilon_p \epsilon_k \sim \epsilon_{p+k}\,.
\ee

A pure lapse transformation 
or a pure dilaton 
 transformation will break condition \eqref{condition}, the correlation
between the number of derivatives and the power of $\epsilon$.  This happens, because
such  redefinitions generate variations from the zero derivative term in the action
and the two-derivative term in the action, and the powers of $\epsilon$ are no longer
correlated with the number of derivatives.

The solution is clear, to leave the correlation between derivatives and 
powers of $\epsilon$ we must do (as we did before) a redefinition of
both the lapse and the dilaton: 
\begin{equation}\label{transformdf}
\Phi = \Phi' + \delta \Phi (\Phi', n') \,, \quad  n = n' + \delta n (\Phi', n') \,, 
\end{equation}
where the variations are related as follows 
\begin{equation}\label{transformnPhi}
\frac{\delta n}{n'} = e^{\delta \Phi} - 1 \,.  \ 
\end{equation}
This is constructed such that $ne^{-\Phi}$ is kept invariant:
\be
ne^{-\Phi} = \bigl( n' + n' (e^{\delta \Phi} -1)\bigr) e^{-\Phi'} e^{-\delta \Phi} 
= n' e^{\delta \phi} e^{-\Phi'} e^{-\delta \Phi } = n' e^{-\Phi'} \,. 
\ee
As a result, the cosmological term does not generate variations, and this
will allow us to  preserve condition~\eqref{condition}.
In the above we take
\be
 \quad \delta \Phi = \sum_{i\geq 1} \epsilon_i F^{(2i)}\,,
\ee 
with $F^{(2 i)}$ generic gauge and duality invariant terms depending on $\hD \Phi$ and containing $2 i$ derivatives, and the $\epsilon_i$ are the constants introduced earlier. 
A completely analogous expression holds for the variation of $n$, with a related 
set of functions $G^{(2i)}$.  

Applying  the variations to the action \eqref{Ibardilaton}, we note that we must
only vary the terms inside the brackets.  
Beginning with the two-derivative 
term and using  the above $\delta\Phi$, each term in 
the variation  will have 
an $\epsilon_i$ accompanied with $2i+2$ derivatives, $2i$ of them from $F^{(2i)}$, and
the other two from the two derivative term being varied.  This is indeed consistent with the structure of the suppression.  
Continuing with the higher-derivative terms, 
varying  $\mathcal{\bar L}^{(2 j + 2)}$
in the action with the $F^{(2k)}$ term of the dilaton variation, 
one gets the product $\epsilon_j \epsilon_k\sim \epsilon_{j+k}$ multiplying  
terms with 
$(2j+ 2k) + 2$ derivatives, which is also consistent with the constraint.  Of course, identical
remarks hold for the variation of $n$.  This shows how that the claimed redefinitions
are consistent with the fall-off conditions.  

Therefore, we can use \eqref{transformnPhi} order by order in $\epsilon$ so to remove terms consistently. This is the procedure developed in \cite{Hohm:2019jgu} for critical strings, with the role of $\alpha'$ played here by the $\epsilon_i$'s satisfying~\eqref{condition}. Using that  logic we can implement field redefinitions as simple substitution rules in the action.  We can in fact conclude that the all-order theory \eqref{Ibardilaton} is totally equivalent to the lowest order one \eqref{ID2}. Indeed, the substitution rule follows
from the equation of motion from part of the theory up to two-derivatives
\begin{equation}\label{dilatonRule}
\bar E_n + \bar E_\Phi= - 2 \hD^2 \Phi  \quad \Rightarrow \quad \hD^2 \Phi \simeq 0 + \mathcal{O}(\epsilon)\,.
\end{equation}
and so we can recursively eliminate any term containing higher derivatives of the 
dilaton (using  integration by parts and dropping total derivatives).

\subsection{Classification of higher derivatives in $D=2$}   
\label{sec_classification} 

The final conclusion of the previous subsection can be easily extended to the
 case when $m$ propagates. To see this, we consider the analogue  to \eqref{Ibardilaton} in the presence of $M$
\begin{equation}\label{Ifull}
\bar I = \int dx \, n\,  e^{-\Phi} \Bigl[ 1 + (\hD\Phi)^2 - \bar M^2 + \sum_{i\geq 1} \epsilon_i \bar{\mathcal{L}}^{(2 i + 2)}(n, \hD \Phi, \bar M) \Bigr]\,,
\end{equation}
where $\bar{\cal L}^{(2i+ 2)}$ contains $2i+2$ derivatives, and the in-built suppression
conditions $\epsilon_i \sim (\epsilon)^i$ holds. We also extended the bar notation to $\bar M = \tfrac1Q M$.
Then, we notice that by extending \eqref{transformnPhi} to
\begin{equation}\label{transformnPhiM}
\begin{aligned}
 \Phi = &\ \Phi' + \delta \Phi(n', \Phi', m') \,, \qquad   n = n' + \delta n (n', \Phi', m')\,, 
\qquad m = m' + \delta m(n', \Phi', m')\,,\\
&\frac{\delta n}{n} = e^{\delta \Phi} - 1\,,  \quad \delta \Phi = \sum_{i\geq 1} \epsilon_i F_{\Phi}^{(2i)}\,, \quad \quad \frac{\delta m}{m} = \sum_{i\geq 1} 
\epsilon_i F_{m}^{(2i)}\,,  
\end{aligned}
\end{equation} 
the in-built suppression feature is satisfied
for the induced terms since transformations of $m$ do not affect the measure 
$ne^{-\Phi}$ and so it remains invariant under the redefinitions \eqref{transformnPhiM}. Finally, in the same way we could use \eqref{dilatonRule} to perform a classification for the dilaton model \eqref{Ibardilaton}, here we can apply the rules
\begin{subequations}\label{Rules}
\begin{align}
E_n + E_\Phi &= 0 \quad \Rightarrow \quad \hD^2 \Phi \simeq \bar M^2 + \mathcal{O}(\epsilon)\,, \label{Rule1}\\
E_m &= 0 \quad \Rightarrow \quad \hD \bar M \simeq \hD \Phi \bar M  + \mathcal{O}(\epsilon)\,, \label{Rule2}
\end{align}
\end{subequations}
where we used the lowest order equations of motion 
given in \eqref{simplifiedequations}. 
The classification of \cite{Hohm:2019jgu} goes through up to the point where redefinitions of the lapse function are needed, which in this case are not allowed because of the extra condition $n' e^{-\Phi'} = n e^{-\Phi}$. 
Specifically, the same step-by-step proof of sec.~2.2 in \cite{Hohm:2019jgu}, 
with the same itemization, proceeds as follows. We assume that to any order in $\epsilon$ any term in 
the action is writable as a product of factors $\hD^k\Phi$ and $\hD^l\bar M$. We can now do field redefinitions of the form \eqref{transformnPhiM}, which in practice consist of applying the rules \eqref{Rules} in the action, in order to establish: 
\begin{itemize}
	\item[(1)] \textit{A factor in an action including $\hD^2\Phi$ can be replaced by a factor with only first derivatives.} \\[0.5ex] 
	This follows directly from \eqref{Rule1}. 
	\item[(2)] \textit{A factor in an action including $\hD \bar M$ can be replaced by a factor with only first derivatives.} \\[0.5ex] 
	This follows directly from \eqref{Rule2}. 
	\item[(3)] \textit{Any action can be reduced so that it only has first derivatives of $\Phi$.} \\[0.5ex]
	The proof proceeds as in \cite{Hohm:2019jgu}: We write any higher derivative as $\hD^{p+2}\Phi=\hD^p(\hD^2\Phi)$, 
	and then integrate by parts the $\hD^p$. Then we substitute $\hD^2\Phi  \rightarrow  \bar M^2$, 
	after which we integrate the derivatives
	 back one-by-one, eliminating any second derivative created, using  (1) or (2).  
	At the end we are left with only first-order derivatives of $\Phi$. 
	\item[(4)] \textit{Any action can be reduced so that it only contains $\bar M$, not its derivatives.} \\[0.5ex]
	The proof is identical to the previous one.
	\item[(5)] \textit{Any higher-derivative term is equivalent to one without any appearance of $\hD\Phi$.} \\[0.5ex] 
	So far we have shown that a generic higher-derivative term in the action is 
	\be
	I = \int d x n e^{-\Phi} (\hD\Phi)^{2p} \bar M^{2l}\;, 
	\ee
	where duality invariance demands even powers of $\bar M$  
	and thus even powers of $\hD\Phi$, since the total number of derivatives
	must be even.  This means that $p, l=0, 1, 2,\ldots$ and $p + l > 1$,
	to have at least four derivatives.
	Using $e^{-\Phi} \hD\Phi = - \hD e^{-\Phi}$ for one of the $\hD\Phi$ factors,
	and then integrating by parts we find
	\be\label{ComputDILL}
	\begin{split}
		I &= - \int d x\,  n \hD(e^{-\Phi}) (\hD\Phi)^{2p-1} \hM^{2l}\,, \\
		&= \int dx\,  n e^{-\Phi} \Big((2p-1)(\hD\Phi)^{2p-2}\hD^2\Phi \hM^{2 l} +(\hD\Phi)^{2p-1} 2 l \hM^{2l-1} \hD \hM\Big)\,, \\
		&\simeq \int dx\,  n e^{-\Phi} \Big((2p-1)(\hD\Phi)^{2p-2} \hM^{2l+2} + 2l (\hD\Phi)^{2p}\hM^{2 l} \Big)\;, 
	\end{split} 
	\ee   
	where we used~(\ref{Rules}) in the last line. The second term in the last line is a multiple of the original term. 
	Thus, bringing it to the left-hand side, 
	\be\label{switchsides}
	(1-2l)I \simeq  (2p-1) \int dx\,  n e^{-\Phi} \,(\hD\Phi)^{2p-2} \hM^{2l+2}\,,  
	\ee
	and so, since $l \neq \frac12$,
	\be\label{dilatondermanip}
	I \simeq \frac{2p-1}{1-2l}  \int dx\,  n e^{-\Phi} \,(\hD\Phi)^{2p-2} \hM^{2l+2}\,.
	\ee 
	Thus, we can systematically reduce the powers of $\hD\Phi$ in steps of two until removing
	all $\hD\Phi$ factors.\footnote{
	Let us justify  \eqref{switchsides}, in which we solve an equivalence relation between actions as if 
	it was  an actual equation. Suppose we established an on-shell equivalence of an 
	action term  $I$ of the form 
	\be\label{onshelleq}
	I \simeq \alpha I +K \;, 
	\ee
	where $\alpha$ is a numerical coefficient and $K$ another action term. In order  to solve  
	$I \simeq \frac{1}{1-\alpha}K$ for  $\alpha\neq 1$ we write 
	\be
	I = (1-\beta)I + \beta I \;, 
	\ee
	where $\beta$ is for now an undetermined parameter. Using (\ref{onshelleq}) in the second term, 
	we have 
	\be
	I \simeq  (1-\beta)I + \beta (\alpha I +K) = (1-\beta(1-\alpha))I+ \beta K \;. 
	\ee
	Since $\beta$ is arbitrary we can choose  $\beta=\frac{1}{1-\alpha}$, so that the terms proportional to $I$ cancel, and 
	$ I \simeq \frac{1}{1-\alpha}K$, as anticipated by solving (\ref{onshelleq}) naively.} 
	
\end{itemize}

The above chain of arguments proved that there is a field basis in which all higher-derivative terms involve only powers of $\hM^2$ and so the most general action is given by
\begin{equation}\label{classification}
I = \int dx \, n\,  e^{-\Phi} \biggl[ 1  
+ (\hD\Phi)^2 - \hM^2 + \sum_{i\geq 1} \epsilon_i \hM^{2i + 2} \biggr]\,, 
\end{equation}
where, if truncated at order $\epsilon^{N - 1}$ we know from \eqref{condition} that the remaining terms are of order $\mathcal{O}(\epsilon^N \hD^{2N + 2})$ and therefore contribute small corrections to the solutions of the truncated theory.

\section{Black hole solutions} \label{blddcljkdfoiw}

This section begins by reviewing the construction of the black hole solution
in the field variables and action with simple duality properties (section~\ref{dimreduvmbb}).   
We point out that the dilaton-lapse theory can
be viewed as giving rise to a zeroth order solution --- a linear dilaton profile, such
that the black hole arises as a perturbation of this solution.   We then begin
to examine the solutions of the general two-dimensional theory with 
spatial dependence only,
using the classification in~(\ref{classification}).   
The general theory is specified by a power series in $\bar{M}$, with just even powers, 
which we call $F(\bar{M})$.  We demonstrate that the general
solution for the fields $m(x)$ and $\Phi(x)$ can be obtained in terms of an integral
involving functions of $\bar{M}$ easily constructed from $F(\bar{M})$.   
The two-derivative black hole
solution, in the gauge $n=1$,  
is then given by $\bar M = \hbox{csch} \, \bar x$ and $e^{\Phi} =  \hbox{csch}\, \bar x$, 
and $m(x)$ can be obtained by suitable integration of $\bar M$. 

The integral formulation is the basis for a perturbative solution of the equations
of motion.  The perturbation is relative to the  two-derivative black hole solution
which is taken to be the zeroth order solution, with the perturbative parameter $\epsilon <1$ that controls the fall-off of the higher-derivative terms in the
action.  We find that at each order of the perturbation the contributions to $\bar M$
and $e^\Phi$ are given by finite polynomials in $\hbox{csch}\, \bar x$.   

We conclude with some analysis of the systematics of non-perturbative solutions,
motivated by the perturbative results.  We use an ansatz where $\bar M$
and $e^\Phi$ are written as series expansions in terms of $\hbox{csch} \, \bar x$ 
and find that the series for $\bar M$ appears to fix both the series for $e^\Phi$ and
for $F(\bar{M})$.    We also discuss the ansatz for an `exact'
solution by Dijkgraaf, Verlinde and Verlinde~\cite{Dijkgraaf:1991ba}.  There is no 
simple method, however, to analyze this solution in our framework 
since in our 
field basis 
the T-duality transformation takes the form $m \to 1/m $, 
which is not the case  in their formulation, 
except in the $k\to \infty$ limit ($k= 9/4$ for the black hole).
More concretely, we see that their ansatz cannot be fit into our formulation.

\subsection{Black holes in the  two-derivative theory} 

For completeness, we begin by re-deriving  the  black hole solution in 
the standard coordinates resembling the Schwarzschild solution in four dimensions. 
The equations of motion are given in~(\ref{simplifiedequations}) and $E_m=0$ implies that $e^{-\Phi} M$ is constant, i.e.,   
 \be\label{Msolution}
  M = e^{\Phi}q\;, \qquad q={\rm const.} 
 \ee
The two remaining equations we put in the equivalent form $E_n=0$, $E_{\Phi}+E_n=0$, 
and write out
the covariant derivatives: 
\be\label{zeroEq}
  \frac{1}{n^2} (\Phi')^2 - \left(\frac{m'}{mn}\right)^2 = Q^2\,,
\qquad    \frac{1}{n}\left(n^{-1}\Phi'\right)'- \left(\frac{m'}{mn}\right)^2 =0 \;.
\ee
We now set
\be
m\, n = 1 \,. 
\ee   
This constraint can be viewed as a gauge condition. Indeed, $mn$ is a density since the lapse $n$ is a density and $m$ is a scalar.  Instead of  gauge fixing to $n=1$,  which we will do at some point, here we gauge fix to have $mn=1$.   Note that in this gauge 
$M = m'$.  
We now make the ansatz 
\be\label{Ansatz}
   e^{-\Phi} = \frac{m}{f}\;. 
 \ee
This relation is just a convenient way to parameterize the dilaton in terms of a function $f$ to be determined. 
 It implies by differentiation 
  \be\label{DilatonDER}
   \Phi' = -\frac{m'}{m} + \frac{f'}{f} \;. 
  \ee
 The relation (\ref{Msolution})  then implies
   \be
       e^{-\Phi}M = e^{-\Phi}\, m' =  \frac{m}{f}m' = q = {\rm const.} \,, 
  \ee
and hence 
 \be\label{mderone}
  mm' = \tfrac{1}{2}(m^2)' = q f \qquad \Rightarrow \qquad 
  m m'' + (m')^2 = q f'
  \;.
 \ee  
Therefore, given $f$, $m$ can be determined by integration: 
 \be\label{msquaredandF} 
  m^2 = 2q F\;, \qquad \text{where}\qquad F' = f  \;. 
 \ee

We next insert the ansatz (\ref{Ansatz}) into the 
second equation in (\ref{zeroEq}) and obtain 
  \be
  \begin{split}
   0 & = m(m\Phi')' - (m')^2 
    = m\left(-m'+m\frac{f'}{f}\right)' -(m')^2 \\
    &= m\left(-m'' + m'\frac{f'}{f} +m\left(\frac{f'}{f}\right)' \right) - (m')^2 
    = m^2\left(\frac{f'}{f}\right)' \;, 
  \end{split} 
  \ee
 where we used (\ref{DilatonDER}) in the first line and both relations  in (\ref{mderone}) in the second  line.  
 Since $m^2\neq 0$ this equation implies that $\frac{f'}{f}$ is constant, hence 
  \be
   f(x) = e^{\gamma x+\delta}\;,  \qquad \gamma, \delta ={\rm const.} 
  \ee
 Its integral then determines $F$ and hence $m^2$ via (\ref{msquaredandF}): 
  \be
  \label{eor77}
   m^2= \frac{2q}{\gamma} e^{\gamma x+\delta} + c\;, 
  \ee
with $c$ a new integration constant.  
Finally, we turn to the equation of motion of the lapse function $n$, the first  equation in  (\ref{zeroEq}).  
Inserting the above ansatz one finds 
 \be
 \begin{split}   
  Q^2 &= m^2(\Phi')^2-(m')^2  = m^2\left(-\frac{m'}{m}+\frac{f'}{f}\right)^2 -(m')^2\\
  &= -2mm' \frac{f'}{f} +m^2\gamma^2
  = -2qf' +\gamma^2\left( \frac{2q}{\gamma} e^{\gamma x+\delta} + c\right)
  = \gamma^2 c\;. 
 \end{split}
 \ee
Thus, the equation just places a relation between the constants in the problem.

Imposing the boundary condition that the metric approaches the Minkowski metric 
far away from the black hole, which we will choose to corresponds to $x\rightarrow -\infty$, 
implies  that $c=1$.  
Then picking  $\gamma=Q$ for $Q$ positive  we have 
from~(\ref{eor77}) and~(\ref{Ansatz}): 
 \be
 \label{30493}
  m^2=1+\frac{2q}{Q} e^{Qx+\delta} \;, \qquad e^{-\Phi} = 
  m\,  e^{-Qx-\delta} \;.     
 \ee
 The second relation here makes it clear that $m$ is, 
 at all points, 
 a real positive number.   
 The above is the well-known black hole solution of  2D string theory and agrees, for instance, 
 with the form given in \cite{Mandal:1991tz}  for 
 \be
  a\equiv -\frac{2q}{Q} \;, \quad \delta=0\;. 
 \ee 
Summarizing, the black hole metric and dilaton read
 \be
 \label{metanddil}
  ds^2 = -m^2(x) dt^2 +\frac{1}{m^2(x)} dx^2\;,   \ \ \ e^{-\Phi} =  m(x) e^{-Qx} \,, 
 \ee
where  
\be
\label{value-m}
m^2(x) =  1 - a e^{Qx} \,,   \qquad    a > 0 \,, \ \ Q > 0 \,. 
\ee

\subsubsection*{Comments on black hole solution} 
Let us close with some brief comments on this BH solution. 
We first note that only if $a > 0$ does $m^2(x)$ vanish for some value of $x$, as we would expect
for a genuine black hole at the event horizon.  
Since $m^2 \to 1$ as $x \to -\infty$ we see that the latter is indeed the asymptotically flat region.   
For the dilaton we have
\be
\label{dil-answer}
\Phi = Q x  - \tfrac{1}{2} \ln m^2 = Qx  - \tfrac{1}{2}  \ln \bigl|  1 - a e^{Qx} \bigr| \,. 
\ee 
Absolute values are needed here for the region beyond the horizon, where $m^2 < 0$.   The need for absolute values also follows because $m$ must be defined as
real and positive, as we mentioned above.   
In the region $x \to -\infty$ we have
\be
\label{dilfaraway}
\Phi = Qx + \tfrac{1}{2} a e^{Qx}  + \tfrac{1}{4} a^2 e^{2Qx} + {\cal O} (e^{3Qx})\,. 
\ee
Here $\Phi \to -\infty$ in the asymptotically flat region, which corresponds to weak coupling.

\medskip
\noindent
\textit{Horizon, singularity, and duality} \\
The Ricci scalar (\ref{CURVATURE}) encodes in 2D the full Riemann curvature.
With $mn=1$ it yields for the above metric
 \be\label{BHcurvature}
  R= -(m^2(x))'' = aQ^2 e^{Qx}  \;. 
 \ee
The metric is singular at the zero of $m^2(x)$, the location of the horizon: 
 \be
  \hbox{Horizon:} \ \ \ \ m^2(x_0)=0 \quad\to \quad   a e^{Qx_0} = 1 \ \ \to \quad x_0=-\tfrac{1}{Q}\ln a\,,  
 \ee
and in analogy to the Schwarzschild solution in 4D we expect this to be a coordinate singularity. Indeed,  the curvature (\ref{BHcurvature}) at this point is regular
 \be\label{Reventhorizon} 
 \hbox{Curvature at horizon:} \ \ \  R(x_0) = Q^2\,.
 \ee 
The black hole singularity is at the point where the curvature $R$  diverges, 
namely as $x \to \infty$: 
 \be\label{BHsingularity} 
  \text{BH singularity:} \qquad x\rightarrow \infty\,. 
 \ee

According to~\cite{Giveon:1991sy} a duality transformation exchanges the horizon of the above black hole with the singularity, 
which we confirm here.  The T-duality transformation keeps $n$ invariant while sending $m\rightarrow m^{-1}$.
Therefore, the dual metric obeys
 \be
  ds^2_{\rm dual} = -\frac{1}{m^2(x)} dt^2 + \frac{1}{m^2(x)} dx^2
  = \frac{1}{m^2(x)}\left(- dt^2 +  dx^2\right) \;. 
 \ee 
The new curvature is obtained 
from~(\ref{CURVATURE}) letting $m\to 1/m$ and $n\to 1/m$:
 \be
R_{\rm dual} = (m^2)'' - { (m^2)' (m^2)' \over m^2} \,. 
 \ee
This shows that the zero of $m^2$, the former horizon location,
now corresponds to a curvature singularity: 
 \be
  R_{\rm dual}(x_0)=\infty\;, 
 \ee
all other points having  finite curvature.  The horizon is now the curvature singularity.
 According to~\cite{Giveon:1991sy}  the original BH singularity $x\to \infty$ 
turns into the horizon of the second one.  Evaluating the
curvature explicitly,   
 \be
  R_{\rm dual} (x)  =  -a Q^2 e^{Qx}  + {a Q^2 e^{Qx} \over \Bigl( 1 - {1\over a e^{Qx}}\Bigr) } 
=   {Q^2 \over 1 - {1\over a} e^{-Qx}} \,.
  \ee
Indeed, for $x\to \infty$, the former position of the curvature singularity,
consistent with~(\ref{Reventhorizon}) we find that 
 \be
 \lim_{x\rightarrow\infty} R_{\rm dual}(x)  = Q^2 \,. 
 \ee

\medskip
\noindent  
\textit{Black hole as  deformation of the lapse-dilaton model solution} \\  
 Consider the lapse-dilaton model action: 
 \be
\label{lformaa}
I = \int dx \, n\,  e^{-\Phi} \bigl(  Q^2 +   (D\Phi)^2 \bigr)\,, 
\ee
Following~(\ref{simplifiedequations}), the equations of motion are 
 \be\label{simplifiedequations2}
  \begin{split}
   E_n &= -(D\Phi)^2  + Q^2  = 0\;, \\ 
    E_{\Phi} &= -2D^2\Phi  + (D\Phi)^2 -  Q^2 = 0  \;.
   \end{split}
 \ee 
The equations imply $D\Phi = Q$, up to an irrelevant sign, and $D^2\Phi=0$.
We take $n=1$ and obtain  $\Phi = Qx$:
\be\label{PhiQx}
n= 1 \,,   \ \ \Phi = Qx\,.
\ee
This can be viewed as a zeroth-order solution.  Once we restore the
$M^2$ term to the action, we have the full black hole equations of motion.  
In the asymptotic region $x \to -\infty$
there are 
{\em small} corrections
to the above zeroth order solution that can be written in the form 
\be
\begin{split}
\Phi = & \ Qx  +  \alpha_1 e^{Qx}  + \alpha_2 e^{2Qx}  + \cdots\\
n = & \  1 + \beta_1 e^{Qx}  + \beta_2 e^{2Qx}  + \cdots \\
m = & \  1  + \gamma_1 e^{Qx}  + \gamma_2 e^{2Qx} + \cdots
\end{split} 
\ee  
Working with $mn = 1$ fixes the expansion of $m$ in terms
of that of $n$, and to zeroth order
one has $m=1$.  Moreover $M =m'$.  
We have checked that the equations
of motion of the full theory now reproduce the terms of the solution for
the black hole --- expanded in the asymptotic region (see~(\ref{dilfaraway})).

\subsection{Black holes in the $\alpha'$-corrected theory}

We have already found a canonical presentation for the action following
our classification of possible duality invariant terms up to field redefinitions.
We can therefore examine how the black hole solution changes
when higher-derivative terms are included. 
In order to do so we rewrite the general form \eqref{classification} of the action as
\be
\label{action-general}
I = \int d\bar x \, n \, e^{-\Phi}  \bigl(1 + (\bar D\Phi)^2 + F(\bar M) \bigr) \,,
\ee
where we are using 
\be
\bar x \equiv Q x\,, \ \ \ \bar \partial \equiv \partial_{\bar x} = \tfrac1Q \partial_x ,\ \ 
 \bar D \equiv \tfrac{1}{n} \bar \partial\,,  \ \ \ \ 
 \bar M \equiv \tfrac1n \bar\partial \ln m = \tfrac{1}{Q} M\,.
 \ee
We set the general expansion 
\be
F(\bar M) \equiv \sum_{i=0}^\infty  c_i \epsilon^i \bar M^{2i+2} 
= - \bar M^2 + \ldots \,, \quad c_0 = -1\,,
\ee
where we rewrote the $\epsilon$ dependence in a way that the structure of in-built suppression (which is always assumed) is made manifest. In this case $\epsilon_i = c_i \epsilon^i$ where all coefficients $c_i$ are of order one and so condition \eqref{condition} is satisfied. 

The equations arising from variation of $m$, $n$, and $\Phi$  give 
\be
\label{3colleqns}
\begin{split}
 \bar D \bigl( e^{-\Phi} f(\bar M) \bigr) & \ = 0 \\
 1 - (\bar D\Phi)^2 - g(\bar M) & \ = 0 \, \\
 -2 \bar D^2 \Phi + (\bar D\Phi)^2 - 1 - F(\bar M) & \ = 0 \,.
 \end{split}
\ee 
Here, 
\be
\begin{split}\label{fg}
f(\bar M)  &\equiv 
  \  F'(\bar M) = \sum_{i=0}^\infty  (2i+2) c_i \epsilon^i \bar M^{2i+1} = - 2\bar M + 4 c_1 \epsilon \bar M^3 + \mathcal{O}(\epsilon^2) \,, \\
g(\bar M) & \equiv \sum_{i=0}^\infty  (2i+1) c_i\epsilon^i \bar M^{2i+2}   = - \bar M^2 + 3 c_1 \epsilon \bar M^4 + \mathcal{O}(\epsilon^2)\,,
\end{split}
\ee
with primes denoting derivative with respect to the argument. 
The following relations are easily checked:
\be
g'(\bar M) = \bar M f'(\bar M)  \,, \ \ \  \ g(\bar M) + F(\bar M) = \bar M f(\bar M) \,. 
\ee
Adding the second and third equations in~(\ref{3colleqns}) we get
\be
\bar D^2 \Phi + \tfrac{1}{2} \bar M f(\bar M) = 0 \,,
\ee
which can replace the third equation to find:
\be
\label{3colleqnss}
\begin{split}
 \bar D \bigl( e^{-\Phi} f(\bar M) \bigr) & \ = 0 \\
 1 - (\bar D\Phi)^2 - g(\bar M) & \ = 0 \, \\
 \bar D^2 \Phi + \tfrac{1}{2} \bar M f(\bar M) & \ = 0 \,.
 \end{split}
\ee 
We now define   
\be
\Omega \equiv e^{-\Phi}\, .
\ee
 The first two equations above
are readily rewritten.  The third gives a more complicated equation that
combined with the second simplifies
\be
\label{3colleqnsss}
\begin{split}
	\bar D \bigl( \Omega  f(\bar M) \bigr) & \ = 0\,, \\
	(\bar D\Omega)^2 + (g(\bar M) - 1) \Omega^2 & \ = 0 \, ,\\
	\bar{D}^2 \Omega - (1 + h(\bar M)) \Omega  & \ = 0 \,,
\end{split}
\ee 
where 
\be\label{hM}
h(\bar M) \equiv \tfrac{1}{2} \bar M f(\bar M) - g(\bar M) \,. 
\ee

\medskip
\goodbreak
\noindent
\textit{A non-perturbative approach to solutions.}

\nobreak
We can focus on the first two equations in~(\ref{3colleqnsss}), then 
solve for $\bar D\Omega/\Omega$ in both and equate the result. This is the procedure
used in \cite{Hohm:2019jgu}. We get
\be
{f'(\bar M)\over f(\bar M)} \bar D\bar M = \pm  \sqrt{1 - g(\bar M)} \,.
\ee
Equivalently, we have, with 
$\bar D\bar M =  {1\over n} {d\bar M\over d\bar x}$,  
\be
\label{309riujdl}
{f'(\bar M) \, d \bar M \over f(\bar M) \sqrt{1 - g(\bar M)} } = \pm  n \,  d\bar x \,.
\ee
By integration we have 
\be
\int^{\bar M}  {f'(M)d M \over f(M) \sqrt{1 - g(M)} } =   \pm  \int^{\bar x}  n(x') dx' \ + C  \,. 
\ee
Here $C$ is a constant of integration. 
This equation fixes a relation between a function of $\bar M$ (the left hand side)
and $\int^{\bar x} n(x') dx'$.  We now adopt  the gauge  
\be
n(\bar x) = 1\,,
\ee
in which the integral condition becomes the simpler  
\be
\label{npsolution}
\int^{\bar M}  {f'(M)d M \over f(M) \sqrt{1 - g(M)} } =   \pm (\bar x  -\bar x_0) \,.
\ee
Here $\bar x_0$ is an integration constant. 
If we have a function $W(M)$ such that  
\begin{equation}\label{dW}
d W \equiv  \frac{f'(M)}{f(M) \sqrt{1 - g(M)}} d M\,,
\end{equation}
the general solution to \eqref{npsolution} is given by
\begin{equation}\label{WQ}
W(\bar M) = \pm (\bar x - \bar x_0)\,,
\end{equation}
a relation that can be inverted to determine $\bar M(\bar x)$. With $\bar M = \bar\partial \log m$, 
in the $n=1$ gauge, this determines $m(\bar x)$. The dilaton is then found from
the first equation in~(\ref{3colleqnsss}), 
\begin{equation}\label{efq}
\Omega  f(\bar M) = q\,,
\end{equation}
with $q$ some constant. By the Bianchi identity, the last equation in~(\ref{3colleqnsss}) holds when the first two hold.

\medskip\noindent
\textit{ Application to the standard black hole solution}  

Let us now re-derive the lowest order black hole solution from the general formula~(\ref{npsolution}). For the two-derivative action we have (see \eqref{fg})
\begin{equation}
f(\bar M) =  - 2 \bar M\,, \quad f'(\bar M) = -2\,, \quad g(\bar M) = - \bar M^2\,,
\end{equation}
and so \eqref{dW} takes the form
\begin{equation}\label{W0}
dW = \frac{d \bar M}{\bar M \sqrt{1 + \bar M^2}}\quad \to \quad W(\bar M) = - \text{arcsch}\, \bar M\,. \end{equation}
By inserting this result into \eqref{WQ} and inverting $W(\bar M)$ we end up with
\begin{equation}\label{M0}
\bar M = \text{csch}\, \bar x =   \bar\partial \ln m \,.
\end{equation}
where we chose a suitable sign and fixed $\bar x_0=0$, without loss of generality. 
This is easily integrated and we obtain
\begin{equation}\label{m0}
m(x) =  \tanh\, \tfrac{\bar x}{2}\,.
\end{equation}
The positivity of $m$ requires
$\bar x = Q x \geq 0$, which determines the allowed space region (we always assume $Q>0$).   This is the {\em exterior} region to the black hole, with $x\to \infty$ the asymptotically flat region.  The horizon is at $x=0$.

The dilaton solution is obtained from \eqref{efq}. Using $f(\bar M) =  -2 \bar M$ and \eqref{M0} we get  
\begin{equation}
\label{dil-exteriorsol} 
e^{\Phi(\bar x)} = -\  \tfrac{2}{q}  \, \hbox{csch} \, \bar x \,. 
\end{equation}
Since $\bar x \geq 0$, the constant $q$ must be negative.

The above is a
solution in the $n=1$ gauge. In order to connect with the zeroth order solution \eqref{value-m} we need to perform a coordinate transformation. The metric we considered here takes the form
\be
ds^2 = - m^2 (x) dt^2  + dx^2\,,  \ \ \hbox{with} \ \  m(x ) = \tanh \tfrac{\bar x}{2} \,.
\ee
We now introduce a coordinate $x'(x)$ via
\begin{equation}\label{dx'dx}
d x' = - m(x) d x\,,
\end{equation}
so that the metric would be of the expected form in the gauge $mn = 1$:
\be
ds^2 = - m^2 (x) dt^2  + {dx'^2\over m^2(x)}  \,,\ee
where in this form $m^2(x)$ must be written in terms of $x'$. 
The sign in~(\ref{dx'dx}) was chosen with hindsight, as the asymptotic region of the
$n=1$ solution is at plus infinity, and the one in the $mn=1$ gauge is at minus infinity. 
Equation~(\ref{dx'dx}) can be easily integrated: 
\begin{equation}\label{xx'}
x' = - \tfrac2Q \ln \cosh \tfrac{Q x}{2} + C \quad  \to  \quad \cosh^2 \tfrac{Q x}{2} = \tfrac1a e^{- Q x'}\,, \quad a > 0\,.
\end{equation}
Here $a$ is an integration constant.   As a result,
\begin{equation}
m(x) =  \tanh \tfrac{Q x}{2} = \sqrt{1 -  \cosh^{-2} \tfrac{Q x}{2}} = \sqrt{1 - a e^{Q x'}}\,,
\end{equation}
in exact agreement with~(\ref{value-m}), obtained in the $mn=1$ gauge. 
The dilatons in the two theories also agree, since they are scalars.  
Indeed, from~(\ref{metanddil})
\be
e^{-\Phi(x')} =  m(x') e^{-Qx'}  =  \tanh \tfrac{Qx}{2}  \cdot  a  \cosh^2  \tfrac{Qx}{2} 
= \tfrac{a}{2} \sinh \, Qx\,. 
\ee
This coincides with $e^{-\Phi}$ in~(\ref{dil-exteriorsol}) if we take $q= -a$.  
This is possible since $a>0$ and $q<0$.

\medskip
\noindent
{\bf A curious case.} 
Let's consider a particular non-perturbative case, where the action contains the two-derivative term and one single correction, a four-derivative term with
a fixed coefficient $c_1 \epsilon = \epsilon_1 = -\tfrac{1}{12}$:
\be
F(\bar M) =  -\bar M^2 - {\bar M^4\over 12 } \ \ \to  \  \   
   f(\bar M) = - 2\bar M - \tfrac{1}{3} \bar M^3 \,, \ \ \  g(\bar M) = -\bar M^2 - \tfrac{1}{4} \bar M^4 \,.  
\ee
In this case 
\be
\sqrt{1- g(\bar M)} =   1 + \tfrac{\bar M^2}{2 }   = - \tfrac{1}{2}  f'(\bar M)\,,
\ee
and equation \eqref{npsolution} simplifies notably:
\be
\int^{\bar M} {d M \over M \bigl( 1 + \tfrac{ M^2}{6} \bigr) } = \pm  \bar x \,. 
\ee
Absorbing the constant of integration into a finite 
shift of $\bar x$ we find the solution 
\be\label{aux4}
\bar M^2  =  \frac{6\,e^{\pm 2 \bar x}}{1 -  e^{\pm 2 \bar x}}\,.
\ee
 In order for the solution to describe the exterior region of a black hole with the asymptotically flat region is at $x \to \infty$, we must
have $\bar M \to 0$ as $x\to \infty$ and that requires choosing the bottom sign.
We thus have
\be\label{aux4}
\bar M^2  =  \frac{6\, e^{-2 \bar x}}{1 -  e^{-2 \bar x}} \, =\,  3 (\coth \bar x - 1) 
\, = \, 3 \bigl( \sqrt{1 + \hbox{csch}^2 \bar x} - \, 1 \bigr) \,.
\ee
With $\bar M$ known, using~\eqref{efq} we can read the dilaton profile
\begin{equation}\label{curiousP}
e^{\Phi(\bar x)} = -\tfrac{\sqrt{3}}{q} \sqrt{\coth \bar x - 1} 
\left(\coth \bar x + 1\right) \, = \,   -\tfrac{\sqrt{3}}{q}  \, \hbox{csch}\, \bar x  \
\sqrt{1+ \coth \bar x }\,.  
\end{equation}
Finally, we can determine $m$ from $\bar M = \bar \partial \ln m$ to find
\begin{equation}\label{curiousm}
m(\bar x) =   e^{\int^{\bar x} 
\bar M(x) d x} =   e^{-\sqrt{6}\, \arcsin e^{-\bar x}}\,,
\end{equation}
where the constant of integration has been chosen to have $m\to 1$ as $x \to \infty$. 
This solution is valid down to $x = 0$. While in the two-derivative theory $m(x)$ vanishes at $x=0$, a point identified as the horizon, here $m(x=0) = \exp (- \sqrt{6} \tfrac{\pi}{2} ) \simeq 0.02$.  It may be of interest to investigate more completely this and related solutions.

\medskip\noindent
\subsection{Systematics of perturbative solutions}  

An analytic expression for the non-perturbative $W(\bar M)$ may not exist in general. However, when considering perturbative solutions in $\epsilon$, equation \eqref{dW} becomes a power series in $\epsilon$, where each term is easier to integrate than the non-perturbative $d W$. In this perturbative regime, a systematic approach exists such that solutions to any order in $\epsilon$ can be obtained from lowest order ones. This algorithm takes \eqref{dW} as the starting point and expand it around small $\epsilon$ to get
\begin{equation}\label{dWe}  
dW = {f'(\bar M) d\bar M \over f(\bar M) \sqrt{1- g(\bar M)} } = \  dW_0 + \epsilon dW_1 + \epsilon^2 dW_2 + \mathcal{O}(\epsilon^3)\,.
\end{equation}
Integrating each of these terms we arrive at the perturbative version of \eqref{WQ}, 
\begin{equation}\label{aux1}
W(\bar M) = W_0(\bar M) + \epsilon W_1(\bar M) + \epsilon^2 W_2(\bar M) + \mathcal{O}(\epsilon^3) = - \bar x\,,
\end{equation}
where we pick the minus sign option, and we are working in the $n=1$ gauge. 
From \eqref{W0} we already know that $W_0(\bar M) = - \text{arcsch} \bar M$, which clearly can be inverted. By doing so, \eqref{aux1} becomes
\begin{equation}\label{aux2}
\begin{aligned}
\bar M &= \text{csch}\left(\bar x + \epsilon W_1(\bar M) + \epsilon^2 W_2(\bar M) \right) + \mathcal{O}(\epsilon^3)\\
&= \text{csch}\,\bar x  + \epsilon\, \text{csch}'\bar x \  W_1(\bar M)\\
& \hspace{2.0cm} + \epsilon^2 \left[ \tfrac12 \text{csch}''\bar x\ W_1^2(\bar M) + \text{csch}'\,\bar x\ W_2(\bar M)\right] + \mathcal{O}(\epsilon^3)\,,
\end{aligned}
\end{equation}
where 
$'$ means derivative with respect to the argument. Then, we expand 
\be
\bar M = \bar M_0 + \epsilon \bar M_1 + \epsilon^2 \bar M_2 + \mathcal{O}(\epsilon^3)\,,\ee
 on both sides of the last equality to read the solution order by order in $\epsilon$
\begin{subequations}\label{allM}
\begin{align}
\bar M_0(\bar x) &= \text{csch}\,  \bar x \,, \label{M0bis}\\
\bar M_1(\bar x) &= \text{csch}'\, \bar  x\ \,  W_1(\bar M_0(\bar x))\,, \label{M01}\\
\bar M_2(\bar x) &= \tfrac12 \text{csch}''\bar x\ \,  W_1^2(\bar M_0(\bar x)) + \text{csch}'\, \bar x\ \,  W_2(\bar M_0(\bar x)) + \text{csch}' \bar x\ \,  W_1'(\bar M_0(\bar x)) \bar M_1(\bar x)\,. 
\end{align}
\end{subequations}
We can see that each order $\bar M_i(\bar x)$ is determined from the lowest order ones. From $\bar M = \bar \partial \ln m$ and \eqref{efq} we can get the perturbative solutions for $m(\bar x)$ and $\Phi(\bar x)$.  The resulting solution will be in the $n=1$ gauge but it can be mapped to the standard form by the use of \eqref{dx'dx} with the now-corrected $m(\bar x)$.

Just as a demonstration of how the above 
 algorithm works in practice, we work out the first $\epsilon$ order explicitly. Up to first order we have (see \eqref{fg})
\begin{equation}\label{fg1}
f(\bar M) = - 2 \bar M + 4 c_1 \epsilon \bar M^3\,, \quad f'(\bar M) = -2 + 12 c_1 \epsilon \bar M^2\,, \quad g(\bar M) = - \bar M^2 + 3 c_1 \epsilon \bar M^4\,.
\end{equation}
Inserting these quantities into \eqref{dWe},   
and expanding up to first order in $\epsilon$  we can read
\begin{equation}
\begin{aligned}
dW_0 &= \frac{1}{\bar M \sqrt{1 + \bar M^2}} d\bar M\,,\\
dW_1 &= -\frac{c_1}{2} \frac{\bar M\left(8 + 5 \bar M^2\right)}{ \left(1+\bar M^2\right)^{\tfrac32}} d\bar M \,.
\end{aligned}
\end{equation}
Each order can be integrated independently to obtain
\begin{equation}\label{W01}
\begin{aligned}
W_0(\bar M) &= - \text{arcsch} \bar M\,,\\
W_1(\bar M) &= -\frac{c_1}{2} \frac{2 + 5 \bar M^2}{ \sqrt{1+\bar M^2}}\,.
\end{aligned}
\end{equation}
Finally, by using \eqref{M01} we can read 
$\bar M_1(\bar x)$:     
\begin{equation}
\begin{aligned}
\bar M_1(\bar x) &= \text{csch}'\, \bar x\  W_1(\bar M_0(\bar x)) = \bigl( - \coth\,\bar x  \ \hbox{csch} \, \bar x  \bigr) \Bigl( -\frac{c_1}{2}\, 
\frac{2 + 5 \, \hbox{csch}^2\, \bar x }{ \sqrt{1+\hbox{csch}^2 \, \bar x }}\ \Bigr) \\
&= c_1 \left(  \hbox{csch}\, \bar x  + \tfrac{5}{2} \, \hbox{csch}^3 \, \bar x\right)\,.
\end{aligned}
\end{equation}
All in all, up to order $\epsilon$, $\bar M(\bar x)$ is given by
\begin{equation}\label{Me01}
\bar M(\bar x) = \text{csch}\, \bar x  +  \epsilon \, c_1 \left( \text{csch}\, \bar x + \tfrac52  \,\text{csch}^3\,  \bar x \right)  + \mathcal{O}(\epsilon^2)\,.
\end{equation}
We easily find $m(\bar x)$ from
\begin{equation}
\begin{split} 
m(\bar x) =\, & \   e^{\int^{\bar x} \bar M(x') dx'} = e^{\int^{\bar x} \, \bar M_0(x') dx'} \Bigl(1 + \epsilon \int^{\bar x} \bar M_1(x') dx'\Bigr) + \mathcal{O}(\epsilon^2)\\
= \, & \  \tanh \tfrac{\bar x}{2}  \Bigl(1 + \epsilon \, c_1 \int^{\bar x} \left( \text{csch}\, x' + \tfrac52  \,\text{csch}^3\, x' \right)dx'\Bigr) + \mathcal{O}(\epsilon^2)\,,
\end{split}
\end{equation}
and doing the integral, 
\begin{equation}
m(\bar x) = \tanh \tfrac{\bar x}{2}\, \Bigl( 1 -\tfrac14 \, c_1 \epsilon \, \bigl[\, \ln \tanh \tfrac{\bar x}{2} + 5  \coth \bar x  \ \hbox{csch}\, \bar x \bigr] \Bigr) + \mathcal{O}(\epsilon^2)\,.
\end{equation}
Finally, the dilaton profile comes from combining \eqref{efq}, \eqref{fg1} and \eqref{Me01} and is given by
\begin{equation}\label{ep1}
\Omega^{-1} = e^{\Phi(\bar x)} = \tfrac1q f(\bar M) =-\tfrac{2}{q}\left[\text{csch}\, \bar x  +  c_1 \epsilon\, \left( \text{csch}\, \bar x + \tfrac12  \,\text{csch}^3\,  \bar x \right)\right] + {\cal O} (\epsilon^2)\,. 
\end{equation}

In view of the pattern emerging at order $\epsilon$ in \eqref{Me01} and \eqref{ep1}, it feels natural to ask whether such structure persists perturbatively to all orders. Indeed, by following an inductive procedure, explained briefly in appendix~\ref{appendix}, we confirmed  
 that the following ansatz  
 can be used to solve  
 \eqref{3colleqnsss} to all orders in $\epsilon$:
\begin{subequations}\label{allordersolution}
\begin{align}
\bar M &= \sum_{p \geq 0} \bar M^{(p)} \epsilon^p\,, \quad  \bar M^{(p)} = \sum_{k = 0}^{p} a_{k}^{(p)} \co{2k + 1}\,,\label{Mall}\\
\Omega^{-1} &= \sum_{p \geq 0} \left[\Omega^{-1}\right]^{(p)} \epsilon^p\,, \quad \left[\Omega^{-1}\right]^{(p)} = \sum_{k = 0}^{p} b_{k}^{(p)} \co{2k + 1}\,. \label{ePall}
\end{align}
\end{subequations}
Here $a_k^{(p)}$ and $b_{k}^{(p)}$ are some order-one coefficients determined completely from the $c_i$ coefficients in the action. For instance, from \eqref{Me01} we can read $a_0^{(0)} = 1, \  a_0^{(1)} = c_1, \ a_1^{(1)} = \tfrac52 c_1$.

\subsection{Systematics of non-perturbative solutions} 

Recall that for the two-derivative action the black hole solution took the 
form
\be
\Omega^{-1} = e^\Phi =  - \tfrac{2}{q}\, \hbox{csch}\, \bar x  \;, \qquad 
 \bar M = \hbox{csch} \, \bar x\,.
\ee
Of course, associated to $\bar M = \bar \partial \ln m$ we have $m = \tanh {\bar x\over 2}$.   Moreover, in perturbative solutions, we found that both $e^\Phi$ and $\bar{M}$ are
written as infinite series in terms of $\hbox{csch} \, \bar x$, beginning with a linear term.
We therefore consider a generalization that we write as follows:
\be
\label{phiMansatz}
\begin{split}
 e^{\Phi}  = & \ \, \,  p_\Phi \circ \hbox{csch} \,, \\
\bar M = & \ p_M^{-1}  \circ \hbox{csch}\,,
\end{split}
\ee
where we use $\circ$ to denote composition of functions, and both sides of the
equalities are functions of $\bar x$.   We have introduced two polynomials, $p_\Phi$
and $p_M$, with $p_M^{-1}$ the inverse function to $p_M$, so that $p_M \circ p_M^{-1} = I$, with $I$ the identity function.   We write
\be
\label{pphiexp}
\begin{split}
  p_\Phi (u) = & \   a_1 u  + a_2 u^2 + {\cal O} (u^3) \,,  \\
  p_M (u) = & \   b_1 u  + b_2 u^2 +  {\cal O} (u^3) \,. 
\end{split}
\ee
From the equation  $e^{-\Phi} f (\bar M) = q$
we have $f(\bar M) = q  e^{\Phi}$ and therefore, as functions of $\bar x$
\be
f \circ \bar M =  q \, p_\Phi \circ \hbox{csch}  \,.
\ee
Composing 
$\bar M^{-1}$ (the inverse function of $\bar M$) from the right on both sides of the above equation we have
\be
f = q \, p_\Phi \circ \hbox{csch} \circ \bar M^{-1} = q \, p_\Phi \circ \hbox{csch} \circ 
\hbox{arcsch} \circ  p_M \,, 
\ee
and therefore we conclude that 
\be
\label{fintermsps} 
f =  q \, p_\Phi  \circ  p_M \,.  
\ee
We conventionally take the normalization of the $m$ field to be determined in the
action by $F(\bar M) = - \bar M^2 + {\cal O} (\bar M^4)$.   Therefore we have
$f(\bar M) = F'(\bar{M}) = -2 \bar M + {\cal O} (\bar M^3)$, and $g(\bar M) 
= - \bar M^2 + {\cal O} (\bar M^4)$.  We thus require that 
\be
f = - 2 I + {\cal O} (I^3) \,,
\ee
This requirement, given the expression for $f$ in~(\ref{fintermsps}) and the expansions in~(\ref{pphiexp}), means the constraints
\be
\begin{split}\label{constraintsf}
q\, a_1 b_1 =  & \, - 2 \,, \\
a_1 b_2 + a_2 \, b_1^2 = & \ \ \  0 \,.
\end{split}
\ee
This means, in particular that $a_1 \not= 0$ and $b_1 \not = 0$. 

We must now calculate $g(\bar{M})$.  
For this we use the middle equation in~(\ref{3colleqnss}) which fixes $g  = 1 - (\bar D \Phi)^2$.   The derivative of $\Phi$ is calculated
from~(\ref{phiMansatz}) and we find
\be
e^\Phi \bar D \Phi = (p'_\Phi \circ \hbox{csch} \, ) 
 \cdot (- \coth \, \cdot\,  \hbox{csch} ) \,.
\ee
Squaring and dividing by $e^{2\Phi}$ gives
\be
(\bar D \Phi)^2 ={ (I^2 \circ p'_\Phi \circ \hbox{csch} \, )  \cdot (  \hbox{csch}^2  +   \hbox{csch}^4 )\over I^2 \circ p_\Phi \circ \hbox{csch} } \,. 
\ee
To view this equation as a function of $\bar M$ we can use the second equation
in~(\ref{phiMansatz}) to write $\hbox{csch} = p_M (\bar M) $ and thus
\be
(\bar D \Phi)^2 ={ (I^2 \circ p'_\Phi \circ p_M\, )  \cdot ( p_M^2  +   p_M^4) 
\over I^2 \circ p_\Phi \circ p_M } \,. 
\ee
We therefore have
\be
\label{fresforg} 
g(\bar M)   =  1 - { ( p'_\Phi (p_M) )^2 \, p_M^2 \, ( 1 + p_M^2)  \over 
  (p_\Phi (p_M))^2 } \,.   
\ee
Here the right-hand side is evaluated with $p_M (\bar M)$.   It is now possible
to check what constraint this result gives given that $g(\bar M) = - \bar{M}^2 + {\cal O} (\bar M^4)$.  Indeed keeping just linear terms on the polynomials we have
that the above equation gives
\be
\begin{split}
g =\ &  1 -  { (a_1 + 2 a_2 p_M )^2  \over 
  (a_1 + a_2 p_M)^2 } + {\cal O} (\bar M^2)  =  1 -  { (1 + 2 \tfrac{a_2}{a_1} b_1 \bar M)^2  \over 
  (1 + \tfrac{a_2}{a_1} b_1 \bar M )^2 } + {\cal O} (\bar M^2) \,, \\
  = & \  - \tfrac{2a_2 b_1}{a_1} \bar M + {\cal O} (\bar M^2) \,.
\end{split}
\ee
Since this linear term must vanish and both $b_1$ and $a_1$ are non-zero, we conclude
that $a_2 = 0$.  The second equation in~(\ref{constraintsf}) then implies that $b_2=0$.
The quadratic terms in the polynomials vanish.   Our perturbative results also hinted in
this direction.  Therefore, we refine the ansatz in~(\ref{pphiexp}) to read
\be
\label{pphiexpnn}
\begin{split}
  p_\Phi (u) = & \   a_1 u  + a_3 u^3 + a_5 u^5 + {\cal O} (u^7)  \,,  \\
  p_M (u) = & \   b_1 u  + b_3 u^3 +  b_5 u^5 + {\cal O} (u^7) \,. 
\end{split}
\ee
The solution now proceeds by first finding $f(\bar M)$ using~(\ref{fintermsps}).
Then, using $g'= M f'$ one finds the associated $g(\bar M)$.
Finally, this result is compared with
the result for $g(\bar{M})$ from~(\ref{fresforg}).  We have found by solving this
system on a computer that the series
$p_M$, defining the metric $\bar M(x)$, fixes the other polynomial $p_\Phi$ as well
as $F(\bar{M})$, the action.  We get, for example, 
\be
a_1 = -{2\over b_1 q}  \,, \ \ \ a_3 = {b_1^2 -1 \over 2 b_1^3 q } \,,  \ \ \   a_5 = {4b_3 -8b_1^5 + 11 b_1^3 -3b_1 \over 32 b_1^6 q} \,. 
\ee  
We also find
\be
f( \bar M) = - 2 \bar{M}   + \bigl( \tfrac{b_1^2}{2} - \tfrac{1}{2} - \tfrac{2b_3}{b_1} \bigr) \bar M^3  + \cdots 
\ee

\medskip
\noindent
{\bf The simplest dilaton profile.} 
Suppose the dilaton profile is now fixed to be the one for the black hole in the two-derivative approximation:
\be  
e^\Phi = -\tfrac{2}{q} \, \hbox{csch} \, \bar x \,.      
\ee
In the language of~(\ref{phiMansatz}) this corresponds to the choice 
$p_\Phi =-\tfrac{2}{q}  I$.    
In this situation, we find that 
$f = -2 p_M (\bar M)$  
as it follows from~(\ref{fintermsps}). 
Finally, from~(\ref{fresforg}) one finds $g(\bar M) = - p_M^2 (\bar M)$.
This means that 
 \be
  g(\bar M) = -\tfrac{1}{4} (f(\bar M))^2 \;. 
 \ee
 Taking the derivative with respect to~$\bar M$, we have 
  \be
   g'(\bar M) = -\tfrac{ 1}{2}f(\bar M)   f'(\bar M)\;.   
  \ee
Recall now that by the common origin of $f$ and $g$ from $F$,
 we have $g'=\bar M f'$, and so comparing with the above we conclude that  
 $f(\bar M) = - 2\bar M$, as expected.  
This is the two-derivative theory, with a quadratic $F(\bar M) = - \bar M^2$.

\medskip
\noindent
{\bf On the ansatz of Dijkgraaf, Verlinde, and Verlinde.}
Based on the form of the Virasoro operator $L_0$ in the CFT description
of the black hole background, Dijkgraaf, Verlinde, and Verlinde (DVV) conjectured
that certain metric and dilaton profiles could represent the {\em exact} 
black hole solution in the $\alpha'$ expansion~\cite{Dijkgraaf:1991ba} (section 4.1).  Their solution requires some 
careful translation, for their dilaton $\phi$ multiplies the action as $e^{\phi}$ 
and ours is the duality invariant dilaton.  With this taken into account, their ansatz
for the dilaton is
\be\label{dilatonSOL}
e^{-\Phi} = \sinh \bar x\;. 
\ee
This is actually the dilaton profile of the two-derivative theory (with 
$q=-2$),  
 and as proven above,
 this means that within our framework the only possible solution is that of the two-derivative theory.  The DVV profile cannot be seen as a solution
 of the $\alpha'$ corrected theory in the presentation we have chosen.  As we mentioned
 in the introduction to this section, this result was expected, as the duality transformations in the DVV profile do not correspond to those of our manifestly
 dual formulation.  
 
 In fact, this is not the only complication.   The proposal 
also says that the metric profile takes the form
 \be
  m^2(\bar x) = \frac{1}{\coth^2(\frac{\bar x}{2}) -\frac{2}{k}}\, = {1 \over  {\cosh \bar x + 1 \over \cosh \bar x - 1} - {2\over k} }  \,.
\ee
With this we now compute:
\be
\bar M = \tfrac{1}{2} \bar \partial m^2 (\bar x) =   {\sinh \bar x\over 
(\cosh \bar x - 1)  \Bigl( 1+ {2\over k} + (1-{2\over k} ) \cosh \bar x\Bigr)} \,.
\ee
Given that $\cosh \bar x = \sqrt{1 + \hbox{csch}^2\bar x} \ / \hbox{csch} \bar x$
we quickly find that  
\be
\bar M =   {\hbox{csch } \bar x\over 
(\sqrt{1 + \hbox{csch}^2\bar x} - \hbox{csch } \bar x)  \Bigl( \Bigl(1+  {2\over k}\Bigr) \hbox{csch } \bar x  + (1-{2\over k} ) \sqrt{1 + \hbox{csch}^2\bar x}\Bigr)} \,.
\ee
This has a Taylor series in the variable~csch $\bar x$, as required for our setup.
Moreover the leading term is linear in~csch $\bar x$ as required for $f(\bar M)$ leading term to be linear in $\bar M$. 
In fact, with $\bar M = p_{M}^{-1}  \circ  \hbox{csch}$, the Taylor series is
\be
p_M^{-1} (u)  =  {u\over 1 - \tfrac{2}{k} }  - {4 u^2 \over  k ( 1 - \tfrac{2}{k} )^2}  + {\cal O} (u^3) 
\ =\  9 u - 144 u^2  + {\cal O}(u^3)  \,,  \ \ \hbox{for} \  k = 9/4\,. 
\ee
We then have $p_M(u ) =  \tfrac{1}{9} u + \tfrac{16}{81} u^2 + \cdots$.  This also
violates the present framework, as we showed above that the polynomial $p_M$ cannot
have a quadratic term.  All in all, our framework does not give any evidence that
the DVV ansatz is a solution.    A solution is meaningful if we also have the 
associated equations
of motion. Those are missing in the DVV conjecture.

\section{Bianchi I cosmology}  \label{bianchioneone}

Following the approach of~\cite{Hohm:2019jgu}, 
in this section we go back to strings 
in the critical dimension (so there is no  
cosmological term) and we classify the $\alpha'$ corrections for backgrounds known as Bianchi type-I cosmologies. 
These backgrounds were also studied in \cite{Bernardo:2020nol}, with a focus on non-perturbative $\alpha'$-complete solutions with matter sources.  
Bianchi type-I cosmologies feature a diagonal metric with a priori independent `scale factors'  on the diagonal. 
In the first subsection we define these backgrounds and determine the corresponding two-derivative action and equations 
of motion. In the second subsection  we classify the most general  higher-derivative terms, 
up to field redefinitions, thereby  
arriving at the classification  of $\alpha'$ corrections.

\subsection{Bianchi I ansatz}

Bianchi type-I (BI) cosmologies are given by a homogeneous but generically  anisotropic metric, 
where the $B$-field vanishes:  
\begin{equation}\label{BI_ansatz}
g_{m n}(t) = a_m(t)^2 \delta_{m n}\,, \quad b_{m n}(t) = 0\,.
\end{equation}
Here $m, n = 1, \dots, d$ are internal indices, and the  indices are not summed over. In general, the  $a_m$ are $d$ independent 
scale factors, but we will consider the case where there 
are only $q \leq d$  
different scale factors. We then have groups of $N_i$ scale factors $a_i$ with $i = 1, \dots, q$ such that $\sum_{i=1}^{q} N_i = d$. 
By definition all $N_i$ are non-zero positive integers.  
The case where all scale factors are different is included for  $q=d$ and $N_i = 1$ for all $i$, 
while the fully isotropic case (FRW) is included for 
$q=1$ and $N_1 = d$. For each of these $q$ scale factors $a_i$ we define the corresponding Hubble parameter $H_i$ as follows:   
\begin{equation}
H_i \equiv \frac{\D a_i}{a_i}\,, \quad i=1, \dots, q\,,
\end{equation}
where $\D \equiv \frac{1}{n}\frac{\partial}{\partial t}$ is a covariant derivative under one-dimensional diffeomorphisms.

Following the notation and conventions of\cite{Hohm:2019jgu},  
the generalized metric and its derivative take the form
\begin{equation}\label{cosmogeneMetric} 
{\cal S}_M{}^N = \begin{pmatrix}   
0 & a^2_m \delta_{m n}\\
a^{-2}_m \delta^{m n} & 0
\end{pmatrix}\,, \quad (\D {\cal S})_{M}{}^N = 2  \begin{pmatrix}   
0 & H_m a^2_m \delta_{m n}\\
- H_m a^{-2}_m \delta^{m n} & 0
\end{pmatrix}\,,
\end{equation}
where there is no sum of repeated indices. 
Even powers of $\D  {\cal S}$ are given by 
\begin{equation}
((\D  {\cal S})^{2 k})_{M}{}^N = (-)^k 2^{2 k}  \begin{pmatrix}  
H_m^{2 k} \delta_m{}^n & 0\\
0 & H_m^{2 k} \delta^m{}_n
\end{pmatrix}\,,
\end{equation}
and its trace is given by
\begin{equation}
\Tr \bigl( (\D  {\cal S})^{2 k}\bigr)    
= (-)^k 2^{2 k + 1}\,  \tr{\left(H^{2 k}_m \delta_m{}^n\right)} = (-)^k 2^{2 k + 1} \sum_{i = 1}^{q} N_i H_i^{2 k}\,,
\end{equation}
where we noted that there are only $q$ different directions and each of them is repeated $N_i$ times.  
In particular, for $k=1$ we have
\begin{equation}
\Tr ((\D  {\cal S})^2) = - 8 \sum_{i = 1}^{q} N_i H_i^2\,.
\end{equation}

The two derivative action in the cosmological setting is~\cite{Hohm:2019jgu}:
\be
I = \int dt \, n \, e^{-\Phi} \left[- (\D \Phi)^2 - \tfrac{1}{8} \Tr ( (\D  {\cal S})^2) \right]\,,
\ee
with $n$ the lapse function, $\Phi$ the dilaton, and $ {\cal S}$ the generalized metric
that we evaluated above for the BI setup with scale factors $a_i$.  Thus the BI two-derivative action reads
\begin{equation}\label{I0}
I^{(0)}_{BI} = \int dt \, n \, e^{-\Phi} \left[- (\D \Phi)^2 + \sum_{i = 1}^{q} N_i H_i^2 \right]\,. 
\end{equation}
The equations of motion are given by  
\begin{subequations}\label{EOM0}
\begin{align}
a_i \frac{\delta I^{(0)}_{BI}}{\delta a_i} &= 0 \quad \Rightarrow \quad \D H_i = \D \Phi H_i\,, \quad i=1, \dots, q\,,\\
\frac{\delta I^{(0)}_{BI}}{\delta \Phi} &= 0 \quad \Rightarrow \quad 2\D^2 \Phi = (\D \Phi)^2 +  \sum_{i = 1}^{q} N_i H_i^2\,, \label{EOMPhi}\\
n \frac{\delta I^{(0)}_{BI}}{\delta n} &= 0 \quad \Rightarrow \quad (\D \Phi)^2 = \sum_{i = 1}^{q} N_i H_i^2\,.\label{EOMn}
\end{align}
\end{subequations}
Using \eqref{EOMn} in \eqref{EOMPhi} we get a more useful version of the equations:   
\begin{subequations}\label{Rules0}
\begin{align}
\D H_i &= \D \Phi H_i\,, \quad i=1, \dots, q\,,\\
\D^2 \Phi &= \sum_{i = 1}^{q} N_i H_i^2\,,\\
(\D \Phi)^2 &= \sum_{i = 1}^{q} N_i H_i^2\,. \label{RuleDPhi2}
\end{align}
\end{subequations}
In the context of BI backgrounds, $O(d,d)$ symmetry reduces  to a  $(\mathbb{Z}_2)^q$ invariance under the transformations   
\begin{align}\label{duality}
a_i &\rightarrow a^{-1}_i \quad \Rightarrow \quad H_i \rightarrow - H_i\,, \quad i=1, \dots, q\,, \nn\\
\Phi &\rightarrow \Phi\,, \quad n \rightarrow n\,.
\end{align}
As expected, \eqref{I0} and \eqref{EOM0} are duality invariant. Note that the duality transformations act on each different scale factor individually. Because of this, while $H_1^2$ or $H_2^2$ are duality invariant, terms like $H_1 H_2$ are not. The latter, however, is invariant under a subset of duality transformations, full-factorized T-dualities, which transform all scale factors simultaneously.

\subsection{Classification}

In the context of string low energy effective actions, equation \eqref{I0} is just the leading contribution to the full action, containing an infinite expansion in $\alpha'$, namely
\begin{equation}
I = \sum_{p\geq 0} \alpha'{}^{p} I^{(p)}\,,
\end{equation}
where each order $I^{(p)}$ can contain a sum of different terms
\begin{equation}
I^{(p)} = \sum_{k} c_{p, k}\, I^{(p)}_k\,,
\end{equation}
where the coefficients $c_{p, k}$ are constant real numbers. Our goal now is to use field redefinitions to reduce these higher-order terms to a minimal set of couplings in the spirit of \cite{Hohm:2019jgu}. To this end we will use the lowest-order equations of motion~\eqref{Rules0} simply as substitution rules in the action which we rewrite here for convenience
\begin{subequations}\label{Rules02}
	\begin{align}
	\D H_i &\simeq \D \Phi H_i\,, \quad i=1, \dots, q\,, \label{DH}\\
	\D^2 \Phi &\simeq \sum_{i = 1}^{q} N_i H_i^2\,, \label{D2Phi}\\
	H_q^2 &\simeq \frac{1}{N_q} \biggl((\D \Phi)^2 - \sum_{i = 1}^{q - 1} N_i H_i^2\biggr)\,.\label{Hq}
	\end{align}
\end{subequations}
The way in which we reordered the last rule distinguishes one particular Hubble parameter over the others. The reason of this split is that, in what follows, we will prove that one of the 
scale factors can be completely removed from higher-order terms, appearing only in the two-derivative theory \eqref{I0}. Without loss of generality, we chose the scale factor 
$a_q$.

For BI universes, the first steps of the classification are similar to those applied in \cite{Hohm:2019jgu}. Specifically, the same step-by-step proof of sec.~\ref{sec_classification} 
with the same itemization, proceeds as follows. We assume that to any order in $\alpha'$ any term in the action is writable as a product of factors $\D^k\Phi$ and $\D^l H_i$ with $i=1, \dots, q$. 
We can now do field redefinitions of $a_i$ and the combination of $\Phi$ and $n$ that yields rules \eqref{DH} and \eqref{D2Phi} in order to establish: 
\begin{itemize}
	\item[(1)] \textit{A factor in an action including $\D^2\Phi$ can be replaced by a factor with only first derivatives.} \\[0.5ex] 
	This follows directly from the substitution rule in \eqref{D2Phi}. 
	\item[(2)] \textit{A factor in an action including $\D H_i$ can be replaced by a factor with only first derivatives.} \\[0.5ex] 
	This follows directly from the first substitution rule \eqref{DH}. 
	\item[(3)] \textit{Any action can be reduced so that it only has first derivatives of $\Phi$.} \\[0.5ex]
	The proof proceeds as in \cite{Hohm:2019jgu}: We write any higher derivative as $\D^{p+2}\Phi=\D^p(\D^2\Phi)$, 
	and then integrate by parts the $\D^p$. Then we substitute $\D^2\Phi  \rightarrow  \sum_{i = 1}^{q} N_i H_i^2$, 
	after which we integrate back one-by-one, eliminating any second derivative created, using  (1) or (2).  
	At the end we are left with only first-order derivatives of $\Phi$. 
	\item[(4)] \textit{Any action can be reduced so that it only contains products of $H_i$, not their derivatives.} \\[0.5ex]
	The proof is identical to the previous one.
	\item[(5)] \textit{Any higher-derivative term is equivalent to one without any appearance of $\D\Phi$.} \\[0.5ex]
	Up to this point, any higher-order term in the action is of the form
	\begin{equation}\label{generic4}
	I = \int dt \, n \, e^{-\Phi} (\D \Phi)^p \prod_{i=1}^{q} H_i^{l_i}\,.
	\end{equation}
	For this term to be a higher derivative one, the total number of derivatives
	must be larger or equal to four:   
	\be
	 p + l  \geq 4 \,,  \ \ \ \ \hbox{with}  \ \ \ \  l \equiv \sum_{i=1}^q  l_i \,. 
	\ee
	Because of duality invariance we must have $l_i \in 2 \mathbb{Z}$,   
	and therefore~$l$ is also even. 
	Some $l_i$ could be zero, and the analysis should hold even
	if all $l_i$ are zero.  Since the total number of derivatives  
	must be even, it follows that $p \in 2 \mathbb{Z}$.  So $p \geq 2$. 
	  
	We proceed to show that any appearance of $\D \Phi$ can be removed. To this end, consider the generic term~\eqref{generic4} with the conditions above and manipulate
	it as follows:  
	\begin{equation}
	\begin{split}
	I &= - \int dt \, n \, \D(e^{-\Phi}) (\D\Phi)^{p-1} \prod_{i=1}^{q} H_i^{l_i}  \\
	&= \int dt \, n \, e^{-\Phi} \biggl( (p-1)(\D\Phi)^{p-2}\D^2\Phi \prod_{i=1}^{q} H_i^{l_i} + (\D\Phi)^{p-1} \sum_{j=1}^{q}(l_j H_j^{-1} \D H_j) \prod_{i=1}^{q} H_i^{l_i}\biggr) \\
	&\simeq \int dt \, n \, e^{-\Phi} \biggl( (p-1)(\D\Phi)^{p-2}\sum_{j=1}^{q} (N_j H_j^2) \prod_{i=1}^{q} H_i^{l_i} + l\, (\D\Phi)^{p} \prod_{i=1}^{q} H_i^{l_i}\biggr)\,.\\
	\end{split} 
	\end{equation}
To pass to the second line we integrated by parts, and to pass to the third
line we  
used \eqref{DH} and \eqref{D2Phi}. 
The last term of the third line is in fact proportional 
to~$I$. 
Bringing it to the left-hand side we end up with 
	\begin{equation}\label{Phireduced}
	I \simeq \frac{p-1}{1 -l}\int dt \, n \, e^{-\Phi} (\D\Phi)^{p-2}\sum_{j=1}^{q} (N_j H_j^2) \prod_{i=1}^{q} H_i^{l_i}\,.
	\end{equation}
Since $l$ is even, $1-l \not=0$, and the right-hand side is well defined. 
Using \eqref{Phireduced} recursively we can reduce the number of $\D \Phi$'s in 
steps of two, ending up with zero since $p$ is even.   
The above formula works fine if all $l_i$ are zero; one simply sets $l=0$ and the rightmost product is replaced by one.
		
\end{itemize}
	
The above chain of arguments proved that there is a field basis in which all higher-derivative terms are of the form
	\begin{equation}\label{generic5}
	I = \int dt \, n \, e^{-\Phi} \prod_{i=1}^{q} H_i^{l_i}\,.
	\end{equation}
	Here $l = \sum_i l_i \geq 4$ is still even.   
	Using the three rules \eqref{Rules02} we now show that any appearance of $H_q$ can be removed from these higher-order terms.
	We begin by rewriting $I$ in a convenient way, 
	\begin{equation}
	I = \int dt \, n \, e^{-\Phi} \prod_{i=1}^{q} H_i^{l_i}
	= \int dt \, n \, e^{-\Phi} H_q^2  H_q^{l_q - 2} \prod_{i=1}^{q -1} H_i^{l_i} \;. 
	\end{equation}
	Using~\eqref{Hq} we now have  
	\begin{equation}
	\begin{split}
	I  &\simeq  \frac{1}{N_q} \int dt \, n \, e^{-\Phi}  \Bigl( (\D \Phi)^2 - \sum_{j = 1}^{q - 1} N_j H_j^2\Bigr) H_q^{l_q - 2} \prod_{i=1}^{q -1} H_i^{l_i}\,,\\
	&\simeq  \frac{1}{N_q} \int dt \, n \, e^{-\Phi} \biggl(\frac{1}{3 - l} \sum_{j=1}^{q} N_j H_j^2 - \sum_{j = 1}^{q - 1} N_j H_j^2\biggr)H_q^{l_q - 2} \prod_{i=1}^{q -1} H_i^{l_i} \,,\\
	&=  \frac{1}{N_q} \int dt \, n \, e^{-\Phi} \biggl(\frac{l - 2}{3 - l} \sum_{j=1}^{q - 1} N_j H_j^2 + \frac{1}{3 - l} N_q H_q^2\biggr) H_q^{l_q - 2} \prod_{i=1}^{q -1} H_i^{l_i}  \,,\\
	&= \frac{1}{N_q} \frac{l - 2}{3 - l} \int dt \, n \, e^{-\Phi} H_q^{l_q - 2} \prod_{i=1}^{q -1} H_i^{l_i} \sum_{j=1}^{q - 1} N_j H_j^2 + \frac{1}{3 - l} I\,.\\
	\end{split}
	\end{equation}
In passing to the second line we used \eqref{Phireduced} on the dilaton term with $p=2$
and with $l\to l-2$.
In passing to the third line we separated the $q$th direction from the first sum.
	Finally, in the last line, we recognized the original term \eqref{generic5}. 
	Since $l$ is even, the denominators are different from zero. 
Taking the latter to the left-hand side and since $l \neq 2$, we end up with 
	\begin{equation}
	I = \int dt \, n \, e^{-\Phi} \prod_{i=1}^{q} H_i^{l_i} \simeq - \frac{1}{N_q} \int dt \, n \, e^{-\Phi} H_q^{l_q - 2}\biggl( \prod_{i=1}^{q -1} H_i^{l_i} \biggr) \sum_{j=1}^{q - 1} N_j H_j^2 \,.
	\end{equation}
	This shows that we can reduce by two units the power of $H_q$, at the expense of increasing the powers of the other $H$'s.  
	Using the result recursively, we are able to redefine away any appearance of $H_q$ 
	at  higher orders in $\alpha'$, and its only appearance 
	is in the two-derivative action \eqref{I0}! 
	The most general higher-derivative term is therefore given by
	\begin{equation}\label{generic_final}
	I = \int dt \, n \, e^{-\Phi} \prod_{i=1}^{q - 1} H_i^{l_i}\,.
	\end{equation}
	
	As a corollary of this result, we get the absence of $\alpha'$ corrections for two particular cases of Bianchi Type-I universes: 
	\begin{itemize}
		\item \textbf{FRW:} In this case we have only one independent  scale factor:
		\begin{equation}
		q=1, \quad N_1 = d, \quad a_1(t) \equiv a(t), \quad H_1(t) \equiv H(t)\,.
		\end{equation}
		Since from \eqref{generic_final} we can always remove one Hubble parameter completely, there is a scheme where there are no $\alpha'$ corrections in the 
		action at all.  
	The action to all orders in $\alpha'$ is just given by the two-derivative theory: 
		\begin{equation}\label{IFRW}
		I_{\rm FRW} = \int dt \, n \, e^{-\Phi} \left[- (\D \Phi)^2 + d \cdot 
		H^2\right]\,.   
		\end{equation}
		
		\item \textbf{Isotropic and static directions:}   
		A slightly more general case than FRW corresponds to the case
		\begin{align}
		q &= 2, \quad N_1 \equiv N, \quad N_2 = d - N, \nn\\
		a_1(t) &\equiv a(t), \quad H_1(t) \equiv H(t), \quad  a_2(t) = \text{const.} \quad \Rightarrow \quad H_2 = 0\,.
		\end{align}
		This is one of the simplest anisotropic backgrounds where we have 
		$N$ isotropic directions and $d-N$ static ones. 
		As in FRW, there is only one Hubble parameter, that we can redefine away. From \eqref{generic_final} we see that there are no higher-order corrections at all. In this case the full action is given by the lowest order one:
		\begin{equation}\label{IStatic}
		I_{\rm static} = \int dt \, n \, e^{-\Phi} \left[- (\D \Phi)^2 + N 
		\cdot H^2\right]\,.   
	      \end{equation}
	\end{itemize}

A few comments are in order concerning the FRW case, which appears to be in conflict with \cite{Hohm:2019jgu}. 
There it was shown that in terms of a generic  generalized metric  ${\cal S}$, encoding a general time-dependent spatial metric and $B$-field, 
there is a minimal field basis in terms of which  all $\alpha'$ corrections are of the form 
\be\label{alpha'correctedSaction} 
 I'[{\cal S}]  = \int dt \, n \, e^{-\Phi} \left\{ \alpha' c_{2, 0} \Tr \left((\D  {\cal S})^4\right) +  \alpha'^2  c_{3, 0} \Tr \left((\D  {\cal S})^6\right) +\cdots  \right\}\;, 
\ee	
with the ellipsis denoting higher order single-trace  and multi-trace terms of even powers of $(\D  {\cal S})$, but 
without factors $\Tr \left((\D  {\cal S})^2\right)$. The coefficients $c_{2, 0}$, $c_{3, 0}$, etc.,  \textit{cannot} be changed 
by field redefinitions and hence have an invariant meaning (and are certainly non-zero).   
Specializing (\ref{alpha'correctedSaction}) then to FRW backgrounds with a single scale factor $a(t)$ one obtains 
corrections of (\ref{IFRW}) with higher powers of $H^2$. Depending on the coefficients 
the resulting theory may exhibit, for instance, non-perturbative de Sitter vacua, which are not visible  in (\ref{IFRW}). 
So how is this result consistent with our above statement that for FRW background all higher-derivative corrections are removable 
by field redefinitions? 

To understand the subtlety let us consider the first correction in $\alpha'$ and let us add a term proportional to $\Tr \left((\D  {\cal S})^2\right)$: 
	\begin{equation}
I^{(1)}[{\cal S}] = \int dt \, n \, e^{-\Phi} \left\{ c_{2, 0} \Tr ((\D  {\cal S})^4) + \xi \left[\Tr ((\D  {\cal S})^2)\right]^2 \right\}\;. 
\end{equation}
As recalled above and shown in  \cite{Hohm:2019jgu}, the new term in here can be removed by field redefinitions: 
the coefficient $\xi$ has no invariant meaning and we may choose $\xi=0$, as done in (\ref{alpha'correctedSaction}). 
We can, however, also choose it to be non-zero and adjust it so that for FRW backgrounds it cancels the contribution 
from the single trace term. Specifically, 
 \be
  \xi \equiv - \tfrac{c_{2, 0}}{2 d}\qquad \Rightarrow \qquad I^{(1)}[{\cal S}_{\rm FRW}] = 0 \;,  
 \ee
where ${\cal S}_{\rm FRW}$ denotes the generalized metric (\ref{cosmogeneMetric}) for a single scale factor. 
Thus, there is a field basis also for ${\cal S}$ so that, \textit{when evaluated on FRW backgrounds}, 
the first-order $\alpha'$ corrections disappear. 
Similar remarks apply to all higher-derivative corrections.

So what, in view of the above discussion,  is the fate of potential  non-perturbative de Sitter vacua? 
It must be emphasized that the above manipulations using field redefinitions order by order in $\alpha'$ 
are strictly perturbative. There is no reason why a non-perturbative solution that is visible  in one perturbative scheme 
must also  be  visible  in another perturbative scheme, and one must await a better understanding of 
non-perturbative string theory even in this simple setting. Relatedly, whether a given solution physically exhibits 
the properties of de Sitter space depends on how one probes the spacetime with matter (or, more concretely, with `clocks')
and in which field basis one couples the clock to the background fields.\footnote{We thank Robert Brandenberger 
for discussions on these points.}

\section{Conclusions and Outlook}\label{concleoridk}

The target space description of string theory contains Einstein gravity coupled to matter fields, universally including an antisymmetric tensor 
($B$-field) and a scalar (dilaton), but importantly it also receives an infinite number of higher-derivative corrections. 
These corrections are only meaningful up to field redefinitions and can hence only be determined up to those  redefinitions.  
Classifying the possible higher-derivative terms up to field redefinitions is hence the important first step of any attempts 
to determine these corrections. In this paper we have explored some surprising, and to the best of our knowledge largely unremarked,  
phenomena that arise when considering higher-derivative corrections for particular backgrounds and/or in non-critical dimensions. 
We have shown that for flat FRW backgrounds (i.e., spatially flat and homogeneous backgrounds with a single scale factor) 
\textit{all} $\alpha'$ corrections are on-shell trivial.  More generally, for so-called Bianchi type I backgrounds 
governed  by $q$ scale factors only $q-1$ receive non-trivial higher-derivative corrections. 
Moreover, and perhaps more thought-provokingly, 
we have emphasized  
that for non-critical dimensions there are \textit{no} invariant terms 
other than the cosmological term, as any term with two or more derivatives can be traded for terms with an arbitrary   number of
derivatives, hence invalidating the familiar  setting  of perturbative $\alpha'$ corrections. 
However, assuming that the numerical coefficients governing the terms in the action fall off in such a manner that 
terms with more derivatives are subleading relative to terms with less derivatives one can give a meaningful classification 
of higher-derivative corrections, as displayed in the main text and applied to the black hole solution of 2D string theory.

Our work is relevant to the understanding of effective field theory in gravitational
theories with a cosmological term.  We have argued that, 
although legal, perturbative
redefinitions that generate variations of the order $1/\alpha'$ cosmological term  
do not respect the canonical structure of $\alpha'$ corrections.
Higher derivative terms in gravity 
are subtle in that their physical effect, small in the context
of a derivative expansion, can 
sometimes be large.  For the 2D string black hole, 
we have seen that they are always intrinsically large.  As discussed 
recently by Horowitz 
{\em et.al.~}in \cite{Horowitz:2023xyl} 
for near-extremal black holes,  
higher derivative terms have an unusually strong effect allowing for a very highly curved geometry near the horizon for generic solutions.

It would be important  to extend the research reported here in various directions, for instance: 
\begin{itemize}

\item Based on the results of  \cite{Hohm:2019jgu,Hohm:2019ccp} various promising string cosmology proposals have  
already been explored, 
see for instance 
\cite{Wang:2019mwi,Wang:2019kez,Bernardo:2020nol,Bernardo:2019bkz, Nunez:2020hxx, Bernardo:2020zlc,Bieniek:2023ubx,Codina:2022onm,Wang:2020eln}. 
In view of the results presented here, these should be revisited for models with two or more scale factors, 
in particular with a focus on semi-realistic embeddings into string theory. 

\item While we have not been able to find the exact black hole solution proposed by 
Dijkgraaf, Verlinde, and Verlinde within our classification of higher-derivative corrections, 
there should be   
a different scheme in which it is a solution, as implied 
by \cite{Tseytlin:1991ht} and \cite{Ying:2022xaj}.   
The proposed solution will likely  only be useful  
if the actual theory of which it is a solution can be written down. 

\item The observation that in gravity theories with cosmological constant  apparently any term in the action  but the cosmological term can be 
removed by field redefinitions arguably  deserves further investigation. The most extreme form of this effect can be displayed 
for an action with cosmological constant $\Lambda$ of the form
 \be\label{cosmologicalAction} 
  S =  - 2\Lambda \int d^Dx \sqrt{-g} \,{\cal L}(g)\;, \ \ \ 
  \hbox{with} \qquad   {\cal L} = 1-\frac{1}{2\Lambda} R+\cdots \,,   
 \ee
which describes Einstein-Hilbert gravity with a cosmological constant, plus arbitrary higher-derivative terms implicit in the ellipsis.  
The field redefinition   
 \be\label{weirdredef} 
  g'_{\mu\nu}  =  [{\cal L}(g)] ^{2/D} g_{\mu\nu}
 \ee
maps the purely cosmological constant theory  
$ S' =  - 2\Lambda \int d^Dx \sqrt{-g'}$
into the action $S$ above.\footnote{We thank Ashoke Sen for pointing out this formulation.} 
Of course, the fractional powers in (\ref{weirdredef}) are generally 
problematic, but for functions ${\cal L}$ of the above form 
they can be  defined  as a power series. 

\item  It may be possible to investigate the possible 
field redefinitions of 
gravitational theories with cosmological term by exploring how observables
of such theories are preserved.  The entropy of black holes 
may provide a useful observable for this analysis.  

\item The arguably most exciting prospect of having an `$\alpha'$-complete' theory governing black holes 
would be as a model for how string theory deals with  
 the black hole singularity (see, e.g., \cite{Perry:1993ry}) and  
the information loss paradox. 
Tentative speculations along these lines are 
as old as string theory itself, 
but what has been lacking are methods 
that allow one to write down concrete theories
 in which such questions can be explored in a precise manner. 
The framework presented should get us closer to that goal.

\end{itemize}

\subsection*{Acknowledgements} 

We thank Robert Brandenberger, Atakan Firat,  Diego Marques, 
Massimo Porrati, Ashoke Sen, and Paul Townsend for comments and discussions.

This work  is supported by the European Research Council (ERC) under the European Union's Horizon 2020 research and innovation program (grant agreement No 771862). 
T.~C.~is supported by the Deutsche Forschungsgemeinschaft (DFG, German Research Foundation) - Projektnummer 417533893/GRK2575 ``Rethinking Quantum Field Theory".

\appendix

\section{All-order perturbative solution}\label{appendix}

In this appendix we prove there exist coefficients $a_k^{(p)}$ and $b_{k}^{(p)}$ such that \eqref{allordersolution} solve the all-order equations of motion \eqref{3colleqnsss}. To see this, it is easier to work with the inverse of \eqref{ePall}, which, together with \eqref{Mall}, is given by
\begin{subequations}\label{allordersolution2}
	\begin{align}
	\bar M &= \sum_{p \geq 0} \bar M^{(p)} \epsilon^p\,, \quad  \bar M^{(p)} = \sum_{k = 0}^{p} a_{k}^{(p)} \co{2k + 1}\,,\label{Mall2}\\
	\Omega &= \sum_{p \geq 0} \Omega^{(p)} \epsilon^p\,, \quad \Omega^{(p)} = \sum_{k = 0}^{p} d_{k}^{(p)} \co{2k - 1}\,, \label{ePinvall}
	\end{align}
\end{subequations}
where the new coefficients $d_{k}^{(p)}$ depend on $ b_{k}^{(p)} $. For instance, $d_0^{(0)} = \frac{1}{b_0^{(0)}} = - \tfrac{q}{2}$. 

The proof follows by induction. Assuming \eqref{allordersolution2} holds up to an including order $\epsilon^{p-1}$, we demonstrate that this structure is also a solution of the order $p$ equations of motion, for some particular choice of coefficients. 

To begin with, we take the second equation of \eqref{3colleqnsss} in the $n=1$ gauge, expand it in powers of $\epsilon$, and read the order $p$ equation
\begin{equation}\label{Yp}
\Omega^{(p)}{}'' - \Omega^{(p)} - \left[h(\bar M) \Omega\right]^{(p)} = 0\,.
\end{equation}
In order to get the coefficients $\left[h(\bar M) \Omega\right]^{(p)}$ we recall the definition of $h(\bar M)$ \eqref{hM}, which, after using the definition of $f(\bar M)$ and $g(\bar M)$, reads
\be\label{hM2}
h(\bar M) = \tfrac{1}{2} \bar M f(\bar M) - g(\bar M)  = - \sum_{i \geq 0} i c_i \epsilon^i {\bar M}^{2 i + 2}\,.
\ee
Multiplying this expansion with $\Omega$, and expanding
\begin{equation}
{\bar M}^{2i + 2} \Omega = \sum_{p\geq 0} \left[{\bar M}^{2i + 2} \Omega\right]^{(p)} \epsilon^p\,,
\end{equation}
we can use the Cauchy product identity for product of power series to read
\begin{equation}\label{hM3}
\begin{aligned}
- \left[h(\bar M) \Omega\right]^{(p)} &= \sum_{i=0}^{p} i c_i \left[\bar M^{2 i + 2} \Omega\right]^{(p - i)} = \sum_{i=0}^{p - 1} (i + 1) c_{i + 1} \left[\bar M^{2 i + 4} \Omega\right]^{(p - i - 1)}\,,\\
\end{aligned}
\end{equation}
where for the second equality we used that the $i=0$ term in the series was zero and then we just renamed the index $i \rightarrow i + 1$. Equation \eqref{hM3} shows explicitly that $\left[h(\bar M) \Omega\right]^{(p)}$ depends on solutions to previous orders only, namely $\Omega^{(k)}$ and $\bar M^{(k)}$ with $k < p$, which is a direct consequence of \eqref{hM2} starting at order $\epsilon$. This last step is crucial in our proof since now, to get $\left[\bar M^{2 i + 4} \Omega\right]^{(k)}$ with $k < p$, we can use the expansion \eqref{allordersolution2}! In order to do so we need a preliminary result: consider two fields $A$ and $B$ admitting an $\epsilon$ expansion with coefficients of the form
\begin{equation}
\begin{aligned}
A^{(p)} = \sum_{k = 0}^{p} a_{k}^{(p)} \co{2k}\,, \quad B^{(p)} = \sum_{k = 0}^{p} b_{k}^{(p)} \co{2k}\,.
\end{aligned}
\end{equation}
It can be shown that the product of them also admits an expansion of the same form, namely
\begin{equation}\label{productrule}
[A B]^{(p)} = \sum_{k = 0}^{p} f(a, b)_{k}^{(p)} \co{2k}\,,\\
\end{equation}
where each $f(a, b)_{k}^{(p)}$ depends on $a_i^{(j)}$ and $b_i^{(j)}$ with $i \leq k$ and $j \leq p$. Since now $A B$ has the same expansion as $A$ and $B$, we can repeat the previous step by multiplying $A B$ with another $A$ or $B$ and use \eqref{productrule} for the new product. By repeating this procedure iteratively, we end up with the extended result
\begin{equation}\label{productrule2}
[A^q B^l]^{(p)} = \sum_{k = 0}^{p} f(a, b)_{k}^{(p)} \co{2k}\,,\\
\end{equation}
with $q$ and $l$ integers. Obviously the coefficients $f(a, b)_{k}^{(p)}$ here are not the same as in \eqref{productrule}, but they follow the same convention.

Coming back to our problem, we can use this intermediate result to get $\left[\bar M^{2 i + 4} \Omega\right]^{(j)}$ by noticing that, up to order $p - 1$,  $\frac{\bar M}{\co{}}$ and $\co{}\, \Omega$ has the same structure as $A$ and $B$ above! (As it can be seen from \eqref{allordersolution2}) Then, we can use \eqref{productrule2} directly with 
\begin{equation}
A \rightarrow \frac{\bar M}{\co{}}\,, \quad B = \co{}\, \Omega\,, \quad q = 2 i + 4\,, \quad l = 1\,,
\end{equation}
so to get
\begin{equation}\label{MOcoeff}
[\bar{M}^{2 i + 4} \Omega]^{(j)} = \co{2i + 3}\, \sum_{k = 0}^{j} f(a, d)_{k}^{(j)} \co{2k}\,,
\end{equation}
which is only valid for $j < p$ and. With these coefficients, we can come back to \eqref{hM3} to get
\begin{equation}\label{hM4}
\begin{aligned}
- \left[h(\bar M) \Omega\right]^{(p)} &= \sum_{i=0}^{p - 1} \sum_{k = 0}^{p - i - 1}  (i + 1) c_{i + 1} f(a, d)_{k}^{(p - i - 1)} \co{2(k + i) + 3}\\
&= \sum_{k=0}^{p - 1} g(a, d)_{k}^{(p-1)} \co{2k + 3}\,,\\
\end{aligned}
\end{equation}
where in the second equality we noticed that both sums can be merged into one, upon defining new combinations $g(a, d)_{k}^{(p-1)}$ which depend on coefficients that were determined from previous steps of the inductive procedure. Finally, we can insert this result back into the order-$p$ equation \eqref{Yp} to get
\begin{equation}\label{Yp2}
\Omega^{(p)}{}'' - \Omega^{(p)} + \sum_{k=0}^{p - 1} g(a, d)_{k}^{(p-1)} \co{2k + 3} = 0\,.
\end{equation}
Then, we can propose
\begin{equation}\label{Omegap}
\Omega^{(p)} = \sum_{k = 0}^{p} d_{k}^{(p)} \co{2k - 1} \ \Rightarrow \ \Omega^{(p)}{}'' = \sum_{k = 0}^{p} d_{k}^{(p)} (2 k - 1)^2 \co{2k - 1} + \sum_{k = 0}^{p} d_{k}^{(p)} 2 k\, \co{2k + 1}\,,
\end{equation}
and check whether there exist coefficients $d_{k}^{(p)}$ such that \eqref{Yp2} is satisfied. By inserting \eqref{Omegap} into \eqref{Yp2}, manipulating the limits of the series and renaming indices we arrive at
\begin{equation}
\begin{aligned}
0 = \sum_{k = 0}^{p - 2} &\left[d_{k + 2}^{(p)} 4 (k + 2)(k + 1) +  d_{k + 1}^{(p)} 2 (k + 1) + g(a, d)_k^{(p-1)} \right] \co{2k + 3}\\
\quad + &\left(2 p\, d_p^{(p)} + g(a, d)_{p-1}^{(p-1)}\right)   \co{2p + 1}\,.
\end{aligned}
\end{equation}
By demanding that each term in the sum vanishes, we arrive at a linear system, involving $p$ equations for the $p$ coefficients $d_k^{(p)}$ with $1 \leq k \leq p$, while $d_0^{(p)}$ is unconstrained\footnote{These undetermined coefficients just renormalize the zeroth-order integration constant $q$.}. One can check that this is an independent system with a unique solution and so $d_k^{(p)}$ are completely determined from previous-order coefficients. This concludes the proof for the dilaton solution, and now we move into $\bar M$.

We start from the first equation of \eqref{3colleqnsss}
\begin{equation}\label{aux3}
f(\bar M) = q\, \Omega^{-1}\, \quad \Rightarrow \quad \left[f(\bar M)\right]^{(p)} = q \sum_{k = 0}^{p} b_{k}^{(p)} \co{2k + 1}\,,
\end{equation}
where we inserted \eqref{ePall} to read the order $p$ coefficient. The coefficients $b_{k}^{(p)}$ are all known from inverting $\Omega^{p}$. They depend on the $d_{k}^{(p)}$ that we just determined in our previous step. Using the definition of $f(\bar M)$, we read $\left[f(\bar M)\right]^{(p)}$
\begin{equation}
\begin{aligned}
\left[f(\bar M)\right]^{(p)} &=\sum_{i=0}^{p} 2(i+1)c_i \left[\bar M^{2i + 1}\right]^{(p-i)}=- 2 \bar{M}^{(p)} + \sum_{i=0}^{p-1} 2(i+2)c_{i+1} \left[\bar M^{2i + 3}\right]^{(p-i-1)}\\
&=- 2 \bar{M}^{(p)} + \sum_{k=1}^{p} f(a)_{k-1}^{(p-1)} \co{2 k + 1}
\,,
\end{aligned}
\end{equation}
where in the second equality we separated the term $i=0$ in the sum and renamed indices $i \rightarrow i + 1$. For the third equality we used \eqref{productrule2} (which is possible because $ \left[\bar M^{2i + 3}\right]^{(p-i-1)} $ involves lower-order solutions of the form \eqref{Mall2}), wrote everything as a power series in $\co{}$ and renamed indices one more time. The coefficients $f(a)_k^{(p-1)}$ are fully determined from previous orders and the $c_i$ coefficients. Finally, by inserting this result into \eqref{aux3} we can isolate $\bar M^{(p)}$ to get
\begin{equation}
\bar{M}^{(p)} = -\tfrac{q}{2} \co{} - \tfrac12\sum_{k=1}^{p}\left[q\, b_k^{(p)} - f(a)_{k-1}^{(p-1)}\right] \co{2 k + 1}\,,
\end{equation}
where we can see that, indeed, $\bar M^{(p)}$ has the desired form \eqref{Mall2} with the new coefficients $a_{k}^{(p)}$ completely determined from $d_k^{(p)}$ and $f(a)_{k}^{(p-1)}$.

\end{document}